\documentclass{aa}  

\usepackage{graphicx}
\usepackage{txfonts}
\usepackage{lipsum}
\usepackage{subcaption}         % necessary for continued figures, example in section 3
\usepackage{lscape}             % to rotate a single page table, example in appendix.
\usepackage{placeins}           % useful with \FloatBarrier, to keep 
\usepackage{booktabs}                                
\titlerunning{GRBs as cosmological probes with quasars}
\authorrunning{Wei Hong et al.}

\begin{document}

%%%%%%%%%%%%%%%%%%%%%%%%%%%%%%%%%%%%%%%%
% if you use custom commands in your title,
% ensure to check your title when submitting!
%%%%%%%%%%%%%%%%%%%%%%%%%%%%%%%%%%%%%%%%
\title{Gamma-ray Bursts as Cosmological Probes Calibrated with Strongly Lensed Quasars and Neural Networks}

   %\subtitle{Subtitle}

%%%%%%%%%%%%%%%%%%%%%%%%%%%%%%%%%%%%%%%%
% Please separate each author with the \and command
%
% Please do not include ORCIDs next to author names.
% Only ORCIDs authenticated by individual authors in EDPS
% editorial system will be taken into account.
% ORCIDs included here will be removed.
%%%%%%%%%%%%%%%%%%%%%%%%%%%%%%%%%%%%%%%%

   \author{Wei Hong\inst{1,2,3}
	\and Luca Izzo\inst{1}\fnmsep\thanks{luca.izzo@inaf.it}
	\and Massimo Della Valle\inst{4}
	\and Orlando Luongo \inst{5,6,7,8,9}
	\and Marco Muccino\inst{5,9,10}
	\and Tong-Jie Zhang\inst{2,3}\fnmsep\thanks{tjzhang@bnu.edu.cn}
}

\institute{INAF - Osservatorio Astronomico di Capodimonte, Salita Moiariello 16, I-80131 Napoli, Italy
	\and Institute for Frontiers in Astronomy and Astrophysics, Beijing Normal University, Beijing 102206, China
	\and Department of Astronomy, Beijing Normal University, Beijing 100875, China
	\and INAF - Osservatorio Astronomico di Padova, Vicolo dell'Osservatorio 5, I-35122 Padova, Italy
	\and Universit\`a di Camerino, Via Madonna delle Carceri, Camerino, 62032, Italy
	\and INAF - Osservatorio Astronomico di Brera, Milano, Italy
	\and Istituto Nazionale di Fisica Nucleare, Sezione di Perugia, Perugia, 06123, Italy
	\and Department of Nanoscale Science and Engineering, University at Albany SUNY, Albany, NY 12222, USA    
	\and Al-Farabi Kazakh National University, Al-Farabi av. 71, 050040 Almaty, Kazakhstan
	\and ICRANet, Piazza della Repubblica 10, 65122 Pescara, Italy}

   \date{Received XX XX, 20XX}

% \abstract{}{}{}{}{}
% 5 {} token are mandatory

\abstract
{Gamma-ray bursts (GRBs) can extend the cosmic distance ladder well beyond the redshift range of Type Ia supernovae, but their cosmological use is limited by calibration circularity and possible redshift-dependent selection or evolution effects.}
{We investigate strongly lensed quasars as an external calibration candidate for the Amati and Combo GRB relations and assess the robustness of the resulting cosmological constraints.}
{We reconstruct the luminosity-distance relation from the TDCOSMO strong-lensing distances using an artificial neural network (ANN), with the absolute scale set by the public TDCOSMO lensing-only calibration. Only GRBs within the redshift range covered by the lens sample are used for calibration, so that the baseline analysis relies on interpolation rather than extrapolation. The calibrated relations are then used in a hierarchical Bayesian analysis of flat $\Lambda$CDM. We also consider $w_0w_a$CDM as a consistency test and examine the possible impact of redshift evolution.}
{For flat $\Lambda$CDM, the Amati relation gives $H_0=78.26^{+17.87}_{-16.30}\,\mathrm{km\,s^{-1}\,Mpc^{-1}}$ and $\Omega_m=0.670^{+0.223}_{-0.276}$. For Combo gives $H_0=78.71^{+13.83}_{-11.29}\,\mathrm{km\,s^{-1}\,Mpc^{-1}}$ and $\Omega_m=0.635^{+0.235}_{-0.242}$. Amati is comparatively stable, whereas Combo is more sensitive to redshift-dependent effects. And injection--recovery tests show that an unmodelled positive Combo trend can bias the inferred cosmology toward higher $H_0$ and $\Omega_m$.}
{Strongly lensed quasars combined with ANN reconstruction provide a feasible external calibration route for extending GRB distance measurements to high redshift. The inferred $H_0$ propagates the adopted TDCOSMO absolute scale rather than constituting an independent GRB determination. Current precision is limited mainly by intrinsic GRB scatter, redshift-dependent systematics, and residual strong-lens modelling uncertainties, with the Combo relation showing the greater sensitivity to sample evolution or selection effects.}


   \maketitle
   \nolinenumbers

\section{Introduction}

The standard cosmological model is tested by several precision probes, including the cosmic microwave background (CMB), baryon acoustic oscillations (BAO), and Type Ia supernovae (SNe Ia) \citep{2013PhR...530...87W,2020A&A...641A...6P,2025JCAP...02..021A,2022ApJ...938..110B}. SNe Ia provide well-established luminosity-distance measurements at low and intermediate redshifts \citep{1993ApJ...413L.105P,2022ApJ...938..110B,2022ApJ...938..113S}, but the available samples become progressively sparser beyond $z\sim1$. Future surveys such as Rubin/LSST \citep{2019ApJ...873..111I,2023ApJ...944..212M}, Euclid \citep{2025A&A...697A...1E}, and the Nancy Grace Roman Space Telescope \citep{2021arXiv211103081R} will improve the situation and extend SNe Ia measurements toward higher redshifts. Distance indicators reaching still higher redshift nevertheless remain useful, both for extending the Hubble diagram and for comparing different calibration routes \citep{2022LRR....25....6M}.

Gamma-ray bursts (GRBs) are sufficiently luminous to be detected deep into the high-redshift Universe. Long GRBs have been observed at $z\geq8$ and possibly beyond \citep{2004RvMP...76.1143P,2006RPPh...69.2259M,2009ARA&A..47..567G,2009Natur.461.1254T,2009Natur.461.1258S,2011ApJ...736....7C}. They are not standard candles, however, and their distances are inferred from empirical correlations affected by substantial intrinsic scatter, detector thresholds, and sample-selection effects \citep{2015NewAR..67....1W,2021Galax...9...77L,2021JCAP...09..042K}. GRBs are therefore most useful as an extension of the Hubble diagram beyond the redshift range where conventional distance indicators are abundant.

We consider the Amati and Combo relations. The Amati relation connects the rest-frame spectral peak energy $E_{p,i}$ with the isotropic-equivalent radiated energy $E_{\rm iso}$ \citep{2002A&A...390...81A,2006MNRAS.372..233A,2008MNRAS.391..577A} and has been widely used to construct GRB Hubble diagrams \citep{2007ApJ...660...16S,2011MNRAS.415.3580D,2017A&A...598A.112D}. The Combo relation combines prompt and X-ray afterglow observables by relating the plateau luminosity $L_0$ to $E_{p,i}$, the rest-frame plateau duration $\tau$, and the post-plateau decay index $\alpha_{\rm PL}$ \citep{2013MNRAS.428..729M,2014A&A...565L..10R,2015A&A...582A.115I,2021ApJ...908..181M}. Since the two samples partly overlap, their cosmological constraints are not statistically independent. Their comparison is instead useful for assessing how different observables and sample properties affect the inferred distances.

A long-standing difficulty in GRB cosmology is the circularity problem. The luminosity correlations require distances for calibration, but these distances are often computed from an assumed cosmological model and then used to constrain cosmology itself \citep{2008ApJ...685..354L,2010A&A...519A..73C,2011MNRAS.415.3580D,2019MNRAS.486L..46A}. Several methods have therefore calibrated GRBs against SNe Ia using local regression, smoothing, Gaussian processes, or other non-parametric reconstructions \citep{2006MNRAS.366.1081S,2007MNRAS.380.1573S,2012PhRvD..85l3530S,2012JCAP...06..036S,2023JCAP...02..014H,2023ApJS..266...27Z,2024ApJS..270...23Z}. These approaches reduce the dependence on an assumed background cosmology, but they still inherit the distance scale and systematic uncertainties of the calibrating data. It is therefore useful to examine an independent calibration channel.

In this work, we use time-delay strongly lensed quasars for this purpose. Their distance information comes from the measured time delays together with lens modelling, stellar kinematics, and line-of-sight information \citep{2022LRR....25....6M}. The inferred distances are not fully model independent: they remain sensitive to the lens mass distribution, mass-sheet transformations, stellar dynamics, line-of-sight convergence, and the distance-duality relation used in converting angular-diameter distances to luminosity distances. We therefore refer to the calibration as independent of an assumed background cosmology, rather than model independent. Its absolute distance scale is taken from the publicly available TDCOSMO lensing analysis. Consequently, the GRB constraints obtained below extend the TDCOSMO distance scale to higher redshift; in particular, they should not be interpreted as an independent GRB determination of $H_0$.

The available strong-lensing distances are few and irregularly distributed in redshift, so we reconstruct their distance--redshift relation with an artificial neural network (ANN). The baseline calibration uses only GRBs within the redshift interval covered by the lens data, $0.2283\leq z\leq0.745$, and therefore does not rely on ANN extrapolation. ANN methods have been applied to several cosmological reconstruction problems \citep{2020ApJS..246...13W,2021MNRAS.503.4581L,2023EPJC...83..956D,2023EPJC...83..304G,2025EPJC...85.1005C,2026arXiv260317500H}. Here we additionally require the accepted reconstructions to satisfy $D_L(z)>0$, ${\rm d}D_L/{\rm d}z>0$, and the low-redshift condition $D_L(0)=0$. Since all reconstructed GRB distances are derived from the same lens sample, they are correlated, and the full ANN covariance is propagated into the calibration likelihood.

The reconstructed distances are used to calibrate the low-redshift Amati and Combo samples, while the calibrated relations are applied to the higher-redshift GRBs. We jointly sample the cosmological parameters, the parameters of the GRB correlations, and their intrinsic dispersions in a hierarchical Bayesian analysis. Flat $\Lambda$CDM is taken as the baseline cosmology. We also consider the Chevallier--Polarski--Linder form, $w(z)=w_0+w_a z/(1+z)$ \citep{2001IJMPD..10..213C,2003PhRvL..90i1301L}, only to test whether the conclusions change when the expansion history is allowed greater freedom; the present GRB data are too weak to provide useful constraints on dynamical dark energy.

Possible redshift-dependent effects remain an important limitation of GRB cosmology. Detector thresholds, incomplete follow-up, sample selection, and population evolution can all change the observed GRB distributions with redshift \citep{2009ApJ...694...76B,2015ApJ...800...31D,2015ApJ...806...44P,2018PASP..130e1001D}. This is especially relevant for the Combo relation because identifying the X-ray plateau also depends on afterglow coverage. We therefore examine the dependence of the cosmological results on the calibration boundary, intrinsic-scatter treatment, parameter priors, sample definition, and an additional redshift-dependent term. For Combo, we compare the historical fixed prescription with a fit in which the evolution parameter is allowed to vary, and use injection--recovery tests to determine how an unmodelled redshift trend would affect the inferred cosmological parameters.

The paper is organised as follows. Section~2 describes the strong-lensing and GRB data. Section~3 presents the ANN reconstruction and covariance treatment. Section~4 describes the GRB calibration and cosmological inference. Section~5 gives the main results and systematic tests. Section~6 discusses their implications and limitations, and Sect.~7 summarizes our conclusions.

\section{Distance-ladder construction with strongly lensed quasars}
\label{sec:distance_ladder}

We combine strongly lensed quasars, which provide the external distance calibration, with the GRB samples used for the Amati and Combo Hubble diagrams. We first describe the lensing distances, then their ANN reconstruction and the GRB observables used in the calibration.

\subsection{Strongly lensed quasars from TDCOSMO2025}
\label{subsec:tdcosmo_data}

We use the TDCOSMO2025 time-delay strong-lensing analysis \citep{2025A&A...704A..63T}, which includes eight strongly lensed quasars with measured time delays and stellar-kinematic information for the deflectors. For a lens at redshift $z_{\rm d}$ and a source at $z_{\rm s}$, the time delay between two images $A$ and $B$ is
\begin{equation}
	\Delta t_{AB}=\frac{D_{\Delta t}}{c}\Delta\tau_{AB},
\end{equation}
where $\Delta\tau_{AB}=\tau(\boldsymbol{\theta}_A,\boldsymbol{\beta})-\tau(\boldsymbol{\theta}_B,\boldsymbol{\beta})$ is the Fermat-potential difference, with $\tau(\boldsymbol{\theta},\boldsymbol{\beta})=|\boldsymbol{\theta}-\boldsymbol{\beta}|^2/2-\phi(\boldsymbol{\theta})$. The time-delay distance is
\begin{equation}
	D_{\Delta t}=(1+z_{\rm d})\frac{D_{\rm d}D_{\rm s}}{D_{\rm ds}},
	\label{eq:time_delay_distance}
\end{equation}
where $D_{\rm d}$, $D_{\rm s}$, and $D_{\rm ds}$ are the corresponding angular-diameter distances.

The inferred distance is affected by the mass-sheet degeneracy. Under the transformation
$\kappa_{\lambda}(\boldsymbol{\theta})=\lambda\kappa(\boldsymbol{\theta})+(1-\lambda)$,
the image configuration can remain unchanged while the inferred time-delay distance is rescaled. Following TDCOSMO, we write
$\lambda=(1-\kappa_{\rm ext})\lambda_{\rm int}$, where $\kappa_{\rm ext}$ describes the external convergence from line-of-sight structure and $\lambda_{\rm int}$ the internal mass-sheet transformation. The time-delay constraint then becomes
\begin{equation}
	D_{\Delta t}=\frac{1}{(1-\kappa_{\rm ext})\lambda_{\rm int}}\frac{c\Delta t_{AB}}{\Delta\tau_{AB}}.
	\label{eq:Ddt_from_time_delay}
\end{equation}

Stellar kinematics provides additional information on the mass distribution. Schematically, the line-of-sight stellar velocity dispersion is $\sigma_{\rm los}^{2}=\frac{1-\kappa_{\rm s}}{1-\kappa_{\rm ds}}\frac{D_{\rm s}}{D_{\rm ds}}c^{2}J(\xi_{\rm lens},\beta_{\rm ani},\lambda_{\rm int})$ where $J$ depends on the lens model, stellar orbital anisotropy, observational aperture and seeing, and the internal mass-sheet transformation. For an extended sheet-like perturbation $J(\xi_{\rm lens},\beta_{\rm ani},\lambda_{\rm int})\approx\lambda_{\rm int}J(\xi_{\rm lens},\beta_{\rm ani})$.

Combining the time-delay and stellar-kinematic information gives
\begin{equation}
	D_{\rm d}=\frac{1}{1+z_{\rm d}}\frac{c\Delta t_{AB}}{\Delta\tau_{AB}}\frac{c^2J(\xi_{\rm lens},\beta_{\rm ani},\lambda_{\rm int})}{\lambda_{\rm int}\sigma_{\rm los}^{2}},
	\label{eq:Dd_from_lensing_kinematics}
\end{equation}
or, using Eq.~(\ref{eq:Ddt_from_time_delay}),
\begin{equation}
	D_{\rm d}=D_{\Delta t}(1-\kappa_{\rm ext})\frac{c^2J(\xi_{\rm lens},\beta_{\rm ani},\lambda_{\rm int})}{(1+z_{\rm d})\sigma_{\rm los}^{2}}.
	\label{eq:Dd_from_Ddt_kinematics}
\end{equation}
Thus the deflector distance is constrained jointly by the time delays and stellar kinematics, while the line-of-sight contribution is included through $\kappa_{\rm ext}$.

In practice, we use the TDCOSMO likelihood products rather than distances inferred from a single best fitting lens model. These products include the relevant lens model, stellar-kinematic, and line-of-sight uncertainties. The distances therefore retain the modelling assumptions of the strong-lensing analysis.

To connect the lensing measurements with the GRB luminosity relations, we use the Etherington distance-duality relation
\begin{equation}
	D_L(z_{\rm d})=(1+z_{\rm d})^2D_{\rm d},
	\label{eq:DL_from_Dd}
\end{equation}
and
\begin{equation}
	\mu(z_{\rm d})=5\log_{10}\left[\frac{D_L(z_{\rm d})}{\mathrm{Mpc}}\right]+25.
	\label{eq:mu_from_DL}
\end{equation}

The relative redshift dependence is determined from the individual lens systems, while the common absolute scale is set by the publicly available TDCOSMO lensing-only distance scale. The GRB analysis therefore propagates this scale to higher redshift rather than providing an independent determination of $H_0$. Further details of the lens-distance preprocessing are given in Appendix~\ref{app:tdcosmo_preprocessing}.

\subsection{ANN reconstruction of the distance relation}
\label{subsec:ann_lens_ladder}

The strong-lensing distances are available at only a few irregularly spaced redshifts. We therefore use an ANN to interpolate them and obtain a continuous distance--redshift relation at the GRB calibration redshifts.

The baseline calibration is restricted to $0.2283\leq z\leq0.745$, the interval covered by the lens sample. We adopt $z_{\rm cut}=0.745$ as the upper calibration boundary and examine its dependence separately by varying $z_{\rm cut}$.

For each calibration GRB, the ANN gives $\mu_{\rm ANN}(z_{\rm GRB})$ and the covariance estimated from the accepted ANN realizations. Since all reconstructed distances are obtained from the same lens systems, their uncertainties are correlated. The off-diagonal covariance is therefore retained in the GRB calibration likelihood. Its construction and numerical treatment are described in Sect.~\ref{sec:ann_reconstruction}.

Accepted reconstructions are also required to satisfy $D_L(z)>0,\frac{{\rm d}D_L}{{\rm d}z}>0$ throughout the reconstruction interval. The low-redshift limit is imposed analytically through $D_L(0)=0$ and $D_L(z)\approx(c/H_0)z$ as $z\rightarrow0$; the point $z=0$ is therefore not supplied to the ANN as a finite distance modulus. The network selection, ensemble construction, and covariance tests are presented in Sect.~\ref{sec:ann_reconstruction}.

\subsection{GRB samples}
\label{subsec:grb_data}

We use GRB samples for the Amati and Combo relations. The Amati sample is selected from the parent catalogue by requiring measured redshifts and the Fermi/GBM prompt-emission quantities needed to determine the spectral peak energy and bolometric fluence \citep{2026arXiv260317500H}. Our Amati catalogue contains 106 GRBs and was constructed by cross-matching the C244 GRB list with events having usable Fermi/GBM prompt-emission measurements, followed by the baseline prompt-quality selection adopted in this work \citep{2026arXiv260401980M}. Since the ANN calibration is restricted to the redshift interval directly supported by the strong-lensing data, three events below the minimum lens redshift, $z=0.2283$, are excluded from the baseline analysis. The final Amati sample therefore contains 103 GRBs, including 19 calibration events at $0.2283\leq z\leq0.745$ and 84 cosmological events at $z\geq0.75$. The rest-frame spectral peak energy is
\begin{equation}
	E_{p,i}=E_{p,\rm obs}(1+z),
	\label{eq:Ep_rest}
\end{equation}
and the bolometric fluence is obtained from the $K$-correction
\begin{equation}
	S_{\rm bol}=S_{\rm obs}\frac{\displaystyle\int_{1/(1+z)}^{10^4/(1+z)}E\Phi(E)\,{\rm d}E}{\displaystyle\int_{E_{\min}}^{E_{\max}}E\Phi(E)\,{\rm d}E},
	\label{eq:Sbolo}
\end{equation}
where $\Phi(E)$ is the prompt-emission photon spectrum and $(E_{\min},E_{\max})$ is the detector-frame energy range. We use the Band spectral form where applicable \citep{1993ApJ...413..281B}.

The Combo sample combines prompt-emission and X-ray afterglow measurements \citep{2015A&A...582A.115I,2021ApJ...908..181M,2026arXiv260401980M}. For the Combo analysis, we adopt the updated C244 catalogue of \citet{2026arXiv260401980M}. We use the published and uniformly processed prompt and X-ray afterglow quantities entering the Combo relation, in particular the rest-frame peak energy $E_{p,i}$, plateau flux $F_0$, and the combined time variable $\mathcal{T}_C=\tau/|1+\alpha_{\rm PL}|$. Seven of the 244 catalogue events lie below the minimum redshift supported by the strong-lensing reconstruction and are not used in the baseline calibration or cosmological inference. This leaves 237 Combo GRBs, consisting of 32 calibration events and 205 high-redshift cosmological events. The relevant quantities are $E_{p,i}$, the plateau flux $F_0$, the rest-frame plateau duration $\tau$, and the post-plateau decay index $\alpha_{\rm PL}$. For a luminosity distance $D_L$
\begin{equation}
	L_0=4\pi D_L^2 F_0,
	\label{eq:combo_L0}
\end{equation}
and the time variable entering the Combo relation is
\begin{equation}
	\mathcal{T}_C=\frac{\tau}{|1+\alpha_{\rm PL}|}.
	\label{eq:combo_time_variable}
\end{equation}

For the baseline analysis, GRBs within $0.2283\leq z\leq0.745$ are used for calibration, while those above the calibration boundary form the cosmological sample. Thus the ANN enters the baseline calibration only through interpolation. The effect of changing the boundary is studied through the $z_{\rm cut}$ scans in Appendix~\ref{app:calibration_support}. The GRB samples used in the Amati and Combo are provided as machine-readable electronic Tables~E1 and E2, respectively.

\section{ANN model optimization and distance reconstruction}
\label{sec:ann_reconstruction}

We now describe the ANN reconstruction used to evaluate the strong-lensing distances at the GRB calibration redshifts. The network architecture is selected using approximate Bayesian computation (ABC) together with a risk-function criterion. Details of the ABC statistics and model-selection tests are given in Appendix~\ref{app:ann_model_selection}.

The input lensing catalogue is
\begin{equation}
	\mathcal{D}_{\rm SL}=\left\{z_{{\rm d},i},D_{L,i},\mu_i,\sigma_{\mu,i}\right\}_{i=1}^{N_{\rm SL}},
	\label{eq:ann_input_catalogue}
\end{equation}
where $z_{{\rm d},i}$ is the deflector redshift and $D_{L,i}$ and $\mu_i$ are obtained from the strong-lensing distances described in Sect.~\ref{subsec:tdcosmo_data}. Because the catalogue contains only a small number of irregularly spaced measurements, the ANN is used to interpolate the distance--redshift relation between the lens redshifts.

To reduce the dynamic range of the input and target variables, we define
\begin{equation}
	x=\ln(1+z),
	\quad
	u=\ln\left[\frac{D_L(z)}{z(1+z)}\right],
	\label{eq:ann_transform}
\end{equation}
and train the network to reconstruct $u(x)$. The luminosity distance is recovered as
\begin{equation}
	D_{L,\boldsymbol{\omega}}(z)=z(1+z)\exp\left[u_{\boldsymbol{\omega}}(x)\right],
	\label{eq:ann_DL_mapping}
\end{equation}
where $\boldsymbol{\omega}$ denotes the network parameters. The corresponding distance modulus is
\begin{equation}
	\mu_{\boldsymbol{\omega}}(z)=5\log_{10}\left[\frac{D_{L,\boldsymbol{\omega}}(z)}{\mathrm{Mpc}}\right]+25.
	\label{eq:ann_mu_mapping}
\end{equation}
The physical conditions imposed on the reconstruction are given in Sect.~\ref{subsec:ann_lens_ladder}.

The network parameters are obtained by minimizing the weighted discrepancy between the reconstructed and lensing distance moduli,
\begin{equation}
	\mathcal{L}_{\mu}=\frac{1}{N_{\rm SL}}\sum_{i=1}^{N_{\rm SL}}\left[\frac{\mu_{\boldsymbol{\omega}}(z_{{\rm d},i})-\mu_i}{\sigma_{\mu,i}}\right]^2.
	\label{eq:ann_weighted_loss_mu}
\end{equation}
The transformed variables are rescaled during training and converted back to physical units before the GRB analysis.

We compare candidate architectures with different numbers of hidden layers $N_{\rm hl}$ and neurons per layer $N_{\rm node}$, as well as different activation functions and optimisers. In the ABC analysis, each architecture is evaluated using its goodness of fit, mean prediction error, and approximate log marginal likelihood. The tolerance $\varepsilon$ is progressively reduced until the relative preference among the surviving models becomes stable; the corresponding definitions are listed in Appendix~\ref{app:ann_model_selection}.

\begin{figure}[ht!]
	\centering
	\includegraphics[width=0.9\linewidth]{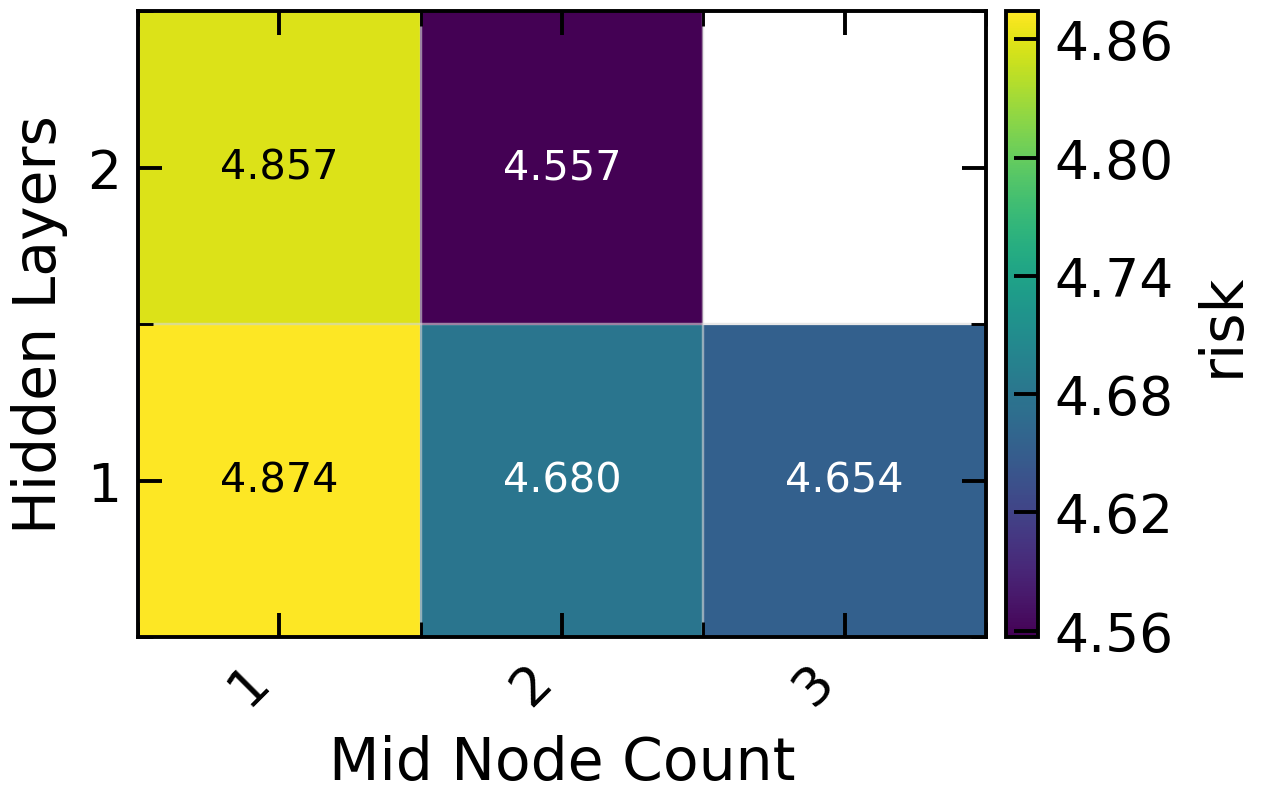}
	\caption{Risk function for the ANN reconstruction. Lower values correspond to a smaller combined contribution from reconstruction bias and variance. The adopted architecture is selected from the ABC-accepted region near the minimum risk.}
	\label{figRHM-lens}
\end{figure}

The accepted architectures are then compared using a risk function that includes reconstruction bias and variance. The adopted network is chosen from the ABC-accepted region close to the minimum risk. Figure~\ref{figRHM-lens} shows the resulting risk landscape. For the selected architecture, we repeat the lens-catalogue realizations and network training to construct an ensemble of reconstructions. The lensing system determines the dependence on redshift, and a common testing method places the reconstruction results on the distance scale of TDCOSMO, which is based solely on the lensing effect.

\begin{figure}[ht!]
	\centering
	\includegraphics[width=1.0\linewidth]{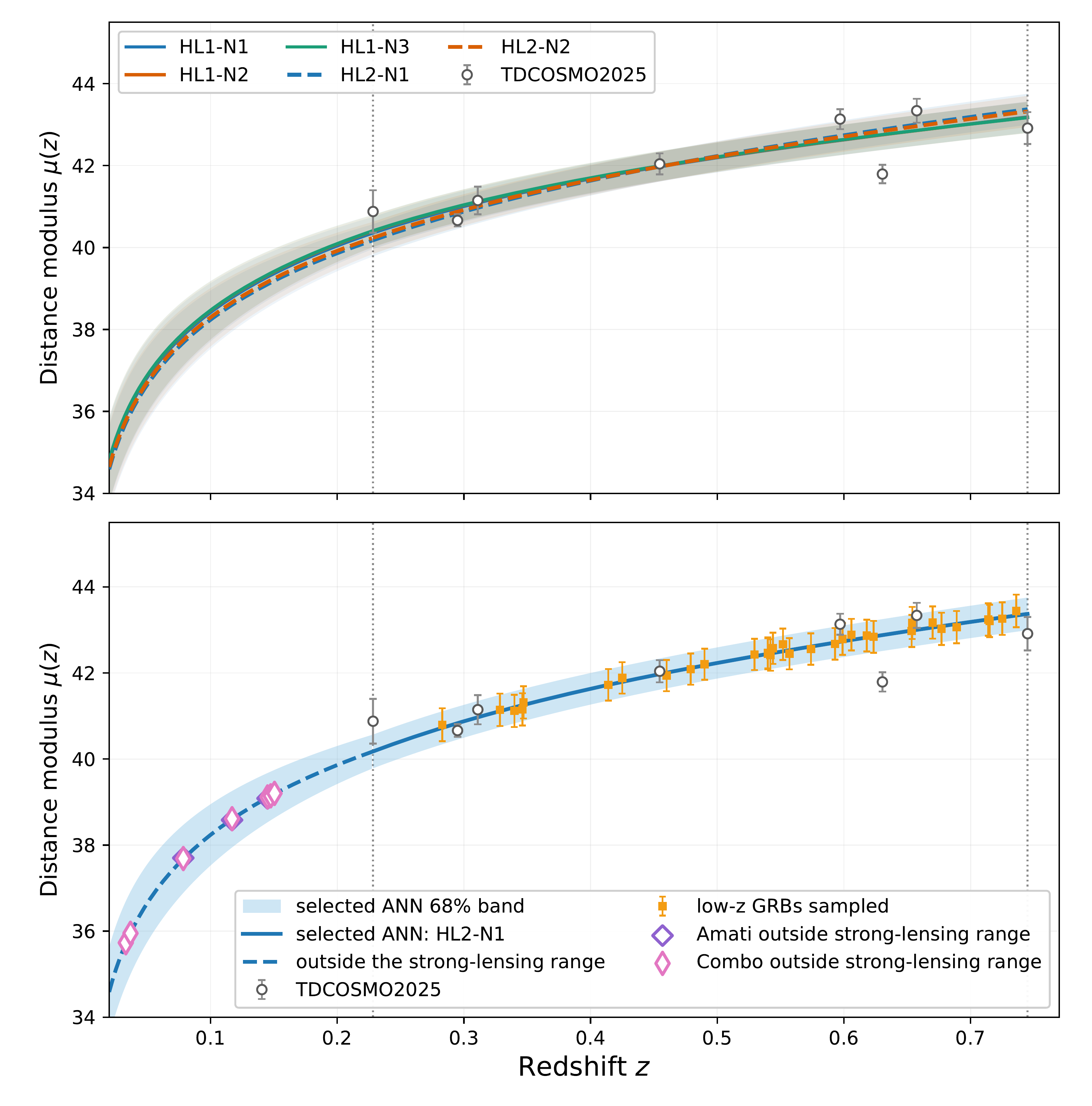}
	\caption{ANN reconstruction of the strong-lensing distance relation. The upper panel shows representative accepted reconstructions, and the lower panel shows the adopted reconstruction at the GRB redshifts. The shaded region gives the marginal $68\%$ uncertainty. GRBs outside $0.2283\leq z\leq0.745$ are shown only for comparison and are not used for calibration.}
	\label{figANN-lens-all}
\end{figure}

Figure~\ref{figANN-lens-all} shows the resulting reconstruction and its evaluation at the GRB redshifts. Only GRBs within $0.2283\leq z\leq0.745$ enter the baseline calibration.

For an ensemble of $N_{\rm acc}$ accepted realizations, the mean reconstructed distance modulus is
\begin{equation}
	\bar{\mu}_{\rm ANN}(z)=\frac{1}{N_{\rm acc}}\sum_{r=1}^{N_{\rm acc}}\mu_r(z),
	\label{eq:ann_mu_mean}
\end{equation}
and the predictive covariance is
\begin{equation}
	C_{\rm ANN}(z_a,z_b)=\frac{1}{N_{\rm acc}-1}\sum_{r=1}^{N_{\rm acc}}\left[\mu_r(z_a)-\bar{\mu}_{\rm ANN}(z_a)\right]\left[\mu_r(z_b)-\bar{\mu}_{\rm ANN}(z_b)\right].
	\label{eq:ann_model_covariance}
\end{equation}
The lens-distance uncertainties are propagated by repeating the reconstruction over realizations of the input catalogue. The resulting covariance therefore includes both the uncertainty of the lensing distances and the variation among the accepted ANN reconstructions. Because the distances at different GRB redshifts are reconstructed from the same lens sample, the off-diagonal terms of $\mathbf{C}_{\rm ANN}$ are retained.

We further include the residual reconstruction dispersion obtained from the ANN validation $\mathbf{C}_{\rm ANN,eff}=\mathbf{C}_{\rm ANN}+\sigma_{\rm rec}^{2}\mathbf{I}$, where $\sigma_{\rm rec}$ accounts for reconstruction variance not represented by the finite ensemble. Its determination is described in Appendix~\ref{app:ann_covariance}. This term is part of the statistical uncertainty and is distinct from the much smaller numerical regularization described below.

Since the reconstructed distances can be strongly correlated, we check the eigenvalue spectrum and condition number of the covariance matrix and require it to remain positive definite. The Gaussian likelihood is evaluated using matrix factorisation and the log-determinant rather than an explicit matrix inverse or determinant. When required for numerical stability, we add a small diagonal term $\mathbf{C}\rightarrow\mathbf{C}+10^{-10}\mathbf{I}$. This term is negligible compared with the statistical variances, and no covariance eigenmode is removed. The eigenvalues, condition number, log-determinant, and numerical stability tests are reported in Appendix~\ref{app:ann_covariance}.

The reconstruction is finally evaluated at each calibration-GRB redshift as
$\mu_{{\rm ANN},j}=\bar{\mu}_{\rm ANN}(z_{{\rm GRB},j})$,
and the corresponding submatrix of $\mathbf{C}_{\rm ANN,eff}$ is used in the calibration likelihood. The Amati and Combo relations are thus calibrated using the same strong-lensing distance reconstruction.

\section{GRB calibration and cosmological inference}
\label{sec:grb_calibration}

We use the ANN-reconstructed strong-lensing distances to calibrate the Amati and Combo relations and infer cosmological parameters from the higher-redshift GRBs. The GRB-relation and cosmological parameters are sampled jointly. The two relations are analysed separately because their samples partly overlap.

\subsection{GRB empirical relations}
\label{subsec:grb_empirical_relations}

The Amati relation connects the rest-frame spectral peak energy $E_{p,i}$ with the isotropic-equivalent radiated energy $E_{\rm iso}$ \citep{2002A&A...390...81A,2006MNRAS.372..233A,2008MNRAS.391..577A}
\begin{equation}
	\log_{10}\left(\frac{E_{{\rm iso},i}}{\rm erg}\right)=A_A+\gamma_A\log_{10}\left(\frac{E_{p,i}}{\rm keV}\right)+\epsilon_{A,i},
	\label{eq:amati_relation}
\end{equation}
where $A_A$ and $\gamma_A$ are the intercept and slope. Since $E_{\rm iso}=\frac{4\pi D_L^2S_{\rm bol}}{1+z}$, the corresponding distance modulus is
\begin{equation}
	\mu_{\rm GRB}^{A}=-97.45+\frac{5}{2}\left[A_A+\gamma_A\log_{10}\left(\frac{E_{p,i}}{\rm keV}\right)-\log_{10}\left(\frac{4\pi S_{\rm bol}}{1+z}\right)\right].
	\label{eq:amati_distance_modulus}
\end{equation}

The Combo relation combines prompt and X-ray afterglow observables
\begin{equation}
	\log_{10}\left(\frac{L_0}{\rm erg~s^{-1}}\right)=A_C+\gamma_C\log_{10}\left(\frac{E_{p,i}}{\rm keV}\right)-\log_{10}\left(\frac{\tau/{\rm s}}{|1+\alpha_{\rm PL}|}\right)+\epsilon_{C,i},
	\label{eq:combo_luminosity_relation}
\end{equation}
where $A_C$ and $\gamma_C$ are the intercept and slope. With $\mathcal{T}_C=\frac{\tau}{|1+\alpha_{\rm PL}|}$ and $L_0=4\pi D_L^2F_0$, we obtain
\begin{equation}
	\begin{aligned}
		\mu_{\rm GRB}^{C}=-97.45+\frac{5}{2}\Bigg[&A_C+\gamma_C\log_{10}\left(\frac{E_{p,i}}{\rm keV}\right)-\log_{10}\left(\frac{\mathcal{T}_C}{\rm s}\right) \\
		&-\log_{10}\left(\frac{F_0}{\rm erg~cm^{-2}~s^{-1}}\right)-\log_{10}(4\pi)\Bigg].
		\label{eq:combo_distance_modulus}
	\end{aligned}
\end{equation}

\subsection{Calibration and cosmological samples}
\label{subsec:cal_cos_samples}

We adopt $z_{\rm cut}=0.745$, corresponding to the upper redshift of the lens-supported interpolation interval. The calibration sample is $\mathcal{D}_{\rm cal}=\left\{i\,\middle|\, 0.2283\leq z_i\leq0.745 \right\}$, while the cosmological sample is $\mathcal{D}_{\rm cos}=\left\{j\,\middle|\,z_j>0.745\right\}$. GRBs below the minimum lens redshift are not used in the baseline calibration.

In this work, the low-redshift calibration sample is used to determine the normalization of the GRB correlations based on the strong-lensing distance scale. The calibrated relations are then applied to the higher-redshift GRBs for cosmological inference. Therefore, the redshift-evolution correction is introduced only for the cosmological sample, while the calibration sample is used to anchor the local correlation. The evolution parameter describes the change of the GRB relation when it is extrapolated to redshifts beyond the range covered by the strong-lensing calibration sample.

For the calibration sample, the reference distance moduli are $\boldsymbol{\mu}_{\rm ANN}=\left\{\mu_{\rm ANN}(z_i)\right\}_{i\in\mathcal{D}_{\rm cal}}$ together with their correlated ANN covariance. The high-redshift GRB distances are instead compared with the theoretical distance modulus of the cosmological model.

We allow different intrinsic dispersions for the calibration and cosmological subsets. For Amati
\begin{equation}
	\left(\sigma_{\mu,A,i}^{s}\right)^2=\left(\frac{5}{2}\right)^2\left[\gamma_A^2\sigma_{\log E_{p,i}}^2+\sigma_{\log S_{{\rm bol},i}}^2+\sigma_{A,s}^2\right], s\in\{{\rm cal},{\rm cos}\},
	\label{eq:sigma_mu_amati}
\end{equation}
and for Combo
\begin{equation}
	\left(\sigma_{\mu,C,i}^{s}\right)^2=\left(\frac{5}{2}\right)^2\left[\gamma_C^2\sigma_{\log E_{p,i}}^2+\sigma_{\log F_{0,i}}^2+\sigma_{\log\mathcal{T}_{C,i}}^2+\sigma_{C,s}^2\right].
	\label{eq:sigma_mu_combo}
\end{equation}

The calibration and cosmological subsets are therefore assigned separate intrinsic-scatter parameters. This allows for the possibility that the dispersion of the high-redshift GRB population differs from that of the calibration sample, for example because of changes in sample composition, measurement quality, or redshift-dependent selection effects. In this way, additional dispersion in the cosmological sample is not absorbed into the low-redshift calibration relation. The corresponding relation-parameter vectors are $\Theta_A=\{A_A,\gamma_A,\sigma_{A,\rm cal},\sigma_{A,\rm cos}\},\Theta_C=\{A_C,\gamma_C,\sigma_{C,\rm cal},\sigma_{C,\rm cos}\}$. The two scatter parameters are sampled independently in the reference analysis. We also repeat the fits imposing $\sigma_{R,\rm cal}=\sigma_{R,\rm cos}$ to examine whether allowing separate dispersions has a significant effect on the cosmological constraints.

\subsection{Hierarchical likelihood}
\label{subsec:hierarchical_likelihood}

For each GRB relation $R\in\{A,C\}$, where $A$ and $C$ denote the Amati and Combo relations, respectively, the total log-likelihood is
\begin{equation}
	\ln\mathcal{L}_{\rm tot}^{R}=\ln\mathcal{L}_{\rm cal}^{R}+\ln\mathcal{L}_{\rm cos}^{R}.
	\label{eq:total_likelihood_R}
\end{equation}
The GRB-relation parameters are common to both terms and are sampled jointly with the cosmological parameters.

For the calibration sample, the distance modulus inferred from the GRB relation is compared with the ANN-reconstructed value at the same redshift. We define the residual as
\begin{equation}
	\Delta_{{\rm cal},i}^{R}=\mu_{{\rm GRB},i}^{R}-\mu_{{\rm ANN},i},
	\quad
	i\in\mathcal{D}_{\rm cal},
	\label{eq:cal_residual_R}
\end{equation}
where $\mu_{{\rm GRB},i}^{R}$ is obtained from the corresponding Amati or Combo relation, and $\mu_{{\rm ANN},i}$ is supplied by the strong-lensing ANN reconstruction. The covariance entering the calibration likelihood is
\begin{equation}
	\mathbf{C}_{\rm cal}^{R}=\mathbf{C}_{\rm ANN}+{\rm diag}\left[\left(\sigma_{\mu,R,i}^{\rm cal}\right)^2\right].
	\label{eq:cal_cov_R}
\end{equation}
The first term contains the correlated uncertainty of the ANN-reconstructed distances, while the diagonal term contains the propagated GRB measurement uncertainty and the intrinsic scatter of the relation. The resulting calibration likelihood is
\begin{equation}
	\ln\mathcal{L}_{\rm cal}^{R}=-\frac{1}{2}\left[\left(\boldsymbol{\Delta}_{\rm cal}^{R}\right)^{\rm T}\left(\mathbf{C}_{\rm cal}^{R}\right)^{-1}
	\boldsymbol{\Delta}_{\rm cal}^{R}+\ln\left|\mathbf{C}_{\rm cal}^{R}\right|+N_{\rm cal}\ln(2\pi)\right],
	\label{eq:cal_likelihood_R}
\end{equation}
where $N_{\rm cal}$ is the number of GRBs in the calibration sample.

For the cosmological sample, the GRB-inferred distance modulus is instead compared directly with the theoretical prediction of the cosmological model,
\begin{equation}
	\Delta_{{\rm cos},j}^{R}=\mu_{{\rm GRB},j}^{R}-\mu_{\rm th}(z_j;\boldsymbol{\theta}),
	\qquad
	j\in\mathcal{D}_{\rm cos},
	\label{eq:cos_residual_R}
\end{equation}
where $\boldsymbol{\theta}$ denotes the cosmological parameters. Since the ANN reconstruction is used only for the calibration sample, the covariance of the cosmological subset contains only the GRB measurement and intrinsic-scatter terms,
\begin{equation}
	\mathbf{C}_{\rm cos}^{R}={\rm diag}\left[\left(\sigma_{\mu,R,j}^{\rm cos}\right)^2\right].
	\label{eq:cos_cov_R}
\end{equation}
The corresponding likelihood is
\begin{equation}
	\ln\mathcal{L}_{\rm cos}^{R}=-\frac{1}{2}\left[\left(\boldsymbol{\Delta}_{\rm cos}^{R}\right)^{\rm T}\left(\mathbf{C}_{\rm cos}^{R}\right)^{-1}
	\boldsymbol{\Delta}_{\rm cos}^{R}+\ln\left|\mathbf{C}_{\rm cos}^{R}\right|+N_{\rm cos}\ln(2\pi)\right],
	\label{eq:cos_likelihood_R}
\end{equation}
where $N_{\rm cos}$ is the number of GRBs entering the cosmological sample.

The two subsets are fitted simultaneously. The calibration likelihood uses the ANN distances rather than the cosmology inferred from the GRBs, while its absolute scale comes from TDCOSMO. The inferred $H_0$ therefore carries this external strong-lensing calibration through the GRB distance ladder.

\subsection{Cosmological models and Bayesian sampling}
\label{subsec:cosmo_models}

For a spatially flat cosmology, the luminosity distance is
\begin{equation}
	D_L(z)=(1+z)\frac{c}{H_0}\int_0^z\frac{{\rm d}z'}{E(z')},
	\label{eq:cosmo_DL}
\end{equation}
with $\mu_{\rm th}(z)=5\log_{10}\left[\frac{D_L(z)}{\rm Mpc}\right]+25$.

We define the dimensionless Hubble parameter as $E(z)\equiv\frac{H(z)}{H_0}$. Therefore, 
for flat $\Lambda$CDM $E(z)=\left[\Omega_m(1+z)^3+1-\Omega_m\right]^{1/2}$, which we take as the baseline cosmological model. We also consider the Chevallier--Polarski--Linder parametrisation $w(z)=w_0+w_a\frac{z}{1+z}$, for which $E(z)=\left[\Omega_m(1+z)^3+(1-\Omega_m)(1+z)^{3(1+w_0+w_a)}\exp\left(-\frac{3w_a z}{1+z}\right)\right]^{1/2}$. The $w_0w_a$CDM runs are used to examine whether the results change when the late-time expansion history is given additional freedom. As shown below, the present GRB data provide only weak constraints on $w_0$ and $w_a$.

We sample the posterior with the affine-invariant ensemble sampler \texttt{emcee} \citep{2010CAMCS...5...65G,2013PASP..125..306F}. The uniform priors used in the reference analyses are listed in Table~\ref{tab:parameter_priors}. The evolution parameters are sampled only in the free-evolution runs; otherwise $\lambda_A=0$ and $\lambda_C=0$ are fixed in the reference Amati and Combo analyses, respectively.
\begin{table}[htbp]
	\centering
	\caption{Uniform priors used in the hierarchical Bayesian analysis.}
	\label{tab:parameter_priors}
	\begin{tabular}{lcc}
		\toprule
		Parameter & Amati & Combo \\
		\midrule
		$H_0\ [{\rm km\,s^{-1}\,Mpc^{-1}}]$& $\mathcal{U}(40,110)$& $\mathcal{U}(40,110)$ \\
		$\Omega_m$& $\mathcal{U}(0.01,0.99)$& $\mathcal{U}(0.01,0.99)$ \\
		$w_0$& $\mathcal{U}(-3,0.5)$& $\mathcal{U}(-3,0.5)$ \\
		$w_a$& $\mathcal{U}(-5,5)$& $\mathcal{U}(-5,5)$ \\
		$A_R$ & $\mathcal U(45,60)$ & $\mathcal U(45,60)$ \\
		$\gamma_R$& $\mathcal U(0.5,2.5)$ & $\mathcal U(0.5,2.5)$ \\
		$\sigma_{R,\rm cal}$& $\mathcal U(0,10)$& $\mathcal U(0,10)$ \\
		$\sigma_{R,\rm cos}$& $\mathcal U(0,10)$& $\mathcal U(0,10)$ \\
		$\lambda_R$& $\mathcal{U}(-0.5,1)$& $\mathcal{U}(-0.5,1)$ \\
		\bottomrule
	\end{tabular}
	\tablefoot{$\mathcal{U}(a,b)$ denotes a uniform prior over $[a,b]$.The parameters $w_0$ and $w_a$ are included only in the$w_0w_a$CDM runs. The $\lambda_R$ priors apply only when the redshift-evolution parameter is allowed to vary
	}
\end{table}

\subsection{Redshift evolution and additional tests}
\label{subsec:robustness_tests}

We introduce redshift-dependent evolution when applying the GRB relations to the cosmological sample. The low-redshift calibration sample is used to determine the normalization of the relations and to set the reference distance scale, while the evolution parameter describes the redshift dependence of the correlations at higher redshifts. For Amati,
\begin{equation}
	S_{\rm bol}^{\prime}=S_{\rm bol}\left[\frac{1+z}{1+z_{\rm ref}}\right]^{-\lambda_A},
	\label{eq:amati_lambda_scaling}
\end{equation}
and for Combo,
\begin{equation}
	F_0^{\prime}=F_0\left[\frac{1+z}{1+z_{\rm ref}}\right]^{-\lambda_C},
	\label{eq:combo_lambda_scaling}
\end{equation}
where $z_{\rm ref}=z_{\rm cut}=0.745$. These transformations shift the inferred distance modulus by
\begin{equation}
	\Delta\mu_R(z)=2.5\lambda_R\log_{10}\left[\frac{1+z}{1+z_{\rm ref}}\right],
	\quad
	R\in\{A,C\},
	\label{eq:lambda_delta_mu}
\end{equation}
while keeping the calibration sample as the reference anchor of the empirical relation.

For Amati, $\lambda_A=0$ defines the reference case, and we also consider fixed-$\lambda_A$ values and a fit in which $\lambda_A$ is free. For Combo, $\lambda_C=0$ is used as the reference, together with the fixed-$\lambda_C$ and a fit in which $\lambda_C$ is sampled jointly with the other parameters.

We further vary the calibration boundary, intrinsic-scatter treatment, cosmological priors, and Combo slope, and examine sample cuts, jackknife removal, bootstrap resampling, and redshift-dependent selection effects. The calibration-boundary scan uses $z_{\rm cut}=0.55,\ 0.60,\ 0.65,\ 0.70,\ 0.745$ all within the redshift range of the lens reconstruction. The Combo injection--recovery test was introduced after the real-data analysis showed a marked dependence on $\lambda_C$. It is therefore not part of the baseline analysis. Its construction and results are given in Sect.~\ref{subsec:lambda_results}.

\section{Results}
\label{sec:results}

We first present the reference cosmological constraints and then examine their dependence on redshift evolution, calibration choices, and sample properties. All production chains satisfy the convergence criterion described in Sect.~\ref{subsec:cosmo_models}. As discussed above, the inferred $H_0$ carries the absolute scale supplied by the TDCOSMO calibration and should not be interpreted as an independent GRB-only measurement.

\subsection{Reference constraints and GRB Hubble diagrams}
\label{subsec:baseline_and_comparison}

Table~\ref{tab:cosmo_results} summarizes the reference hierarchical constraints. The Amati analysis adopts $\lambda_A=0$, same with the Combo reference $\lambda_C=0$. Flat $\Lambda$CDM is taken as the baseline cosmological model.

\begin{table*}[htbp]
	\centering
	\caption{Reference constraints from the hierarchical Amati and Combo analyses. Values are posterior medians with central $68\%$ uncertainties, and central $95\%$ ranges are shown in parentheses.}
	\label{tab:cosmo_results}
\begin{tabular}{lcccc}
	\toprule
	Parameter & Amati flat $\Lambda$CDM & Amati $w_0w_a$CDM & Combo flat $\Lambda$CDM & Combo $w_0w_a$CDM \\
	\midrule
	\addlinespace
	\multicolumn{5}{l}{\textit{Cosmological parameters}} \\
	\cmidrule(r){1-1}\cmidrule(lr){2-2}\cmidrule(lr){3-3}\cmidrule(lr){4-4}\cmidrule(l){5-5}
	$H_0$ & $78.26_{-16.30\,(-28.90)}^{+17.87\,(+29.05)}$ & $76.50_{-17.49\,(-30.39)}^{+19.23\,(+30.72)}$ & $78.71_{-11.29\,(-20.01)}^{+13.83\,(+26.38)}$ & $78.93_{-12.14\,(-21.31)}^{+15.86\,(+27.87)}$ \\
	$\Omega_m$ & $0.670_{-0.276\,(-0.472)}^{+0.223\,(+0.305)}$ & $0.644_{-0.298\,(-0.537)}^{+0.241\,(+0.330)}$ & $0.635_{-0.242\,(-0.395)}^{+0.235\,(+0.336)}$ & $0.630_{-0.250\,(-0.434)}^{+0.241\,(+0.341)}$ \\
	$w_0$ & -- & $-1.414_{-1.085\,(-1.508)}^{+1.226\,(+1.801)}$ & -- & $-1.575_{-0.995\,(-1.358)}^{+1.281\,(+1.943)}$ \\
	$w_a$ & -- & $-0.166_{-3.230\,(-4.578)}^{+3.094\,(+4.775)}$ & -- & $-1.012_{-2.668\,(-3.779)}^{+2.796\,(+5.027)}$ \\
	\addlinespace
	\multicolumn{5}{l}{\textit{GRB-correlation parameters}} \\
	\cmidrule(r){1-1}\cmidrule(lr){2-2}\cmidrule(lr){3-3}\cmidrule(lr){4-4}\cmidrule(l){5-5}
	$A$ & $49.693_{-0.335\,(-0.675)}^{+0.336\,(+0.665)}$ & $49.717_{-0.342\,(-0.681)}^{+0.336\,(+0.663)}$ & $50.001_{-0.175\,(-0.350)}^{+0.177\,(+0.348)}$ & $50.012_{-0.177\,(-0.352)}^{+0.174\,(+0.344)}$ \\
	$\gamma$ & $1.132_{-0.121\,(-0.240)}^{+0.123\,(+0.249)}$ & $1.122_{-0.123\,(-0.241)}^{+0.127\,(+0.251)}$ & $0.616_{-0.067\,(-0.132)}^{+0.067\,(+0.132)}$ & $0.614_{-0.066\,(-0.131)}^{+0.066\,(+0.132)}$ \\
	\addlinespace
	\multicolumn{5}{l}{\textit{Intrinsic scatter parameters}} \\
	\cmidrule(r){1-1}\cmidrule(lr){2-2}\cmidrule(lr){3-3}\cmidrule(lr){4-4}\cmidrule(l){5-5}
	$\sigma_{\rm cal}$ & $0.750_{-0.116\,(-0.206)}^{+0.151\,(+0.352)}$ & $0.751_{-0.117\,(-0.205)}^{+0.152\,(+0.348)}$ & $0.417_{-0.061\,(-0.112)}^{+0.074\,(+0.160)}$ & $0.417_{-0.061\,(-0.111)}^{+0.073\,(+0.161)}$ \\
	$\sigma_{\rm cos}$ & $0.576_{-0.044\,(-0.083)}^{+0.051\,(+0.107)}$ & $0.576_{-0.045\,(-0.083)}^{+0.050\,(+0.106)}$ & $0.345_{-0.021\,(-0.039)}^{+0.022\,(+0.045)}$ & $0.344_{-0.020\,(-0.039)}^{+0.022\,(+0.045)}$ \\
	\bottomrule
\end{tabular}
	
	\vspace{0.4em}
	\footnotesize{
		\textit{Notes.}
		Uncertainties are written as $X_{-a(-c)}^{+b(+d)}$, where $a,b$ and $c,d$ denote the central $68\%$ and $95\%$ credible intervals, respectively.$H_0$ is in ${\rm km\,s^{-1}\,Mpc^{-1}}$.
	}
\end{table*}

The Amati and Combo relations give overlapping flat-$\Lambda$CDM constraints within the present uncertainties. Combo has a smaller effective scatter and therefore somewhat greater statistical weight, although it is also more sensitive to redshift-dependent effects, as discussed below. Because the two GRB samples partly overlap, their constraints should not be regarded as statistically independent.

Allowing the expansion history to vary within $w_0w_a$CDM broadens the cosmological posteriors substantially. The $w_0$ and $w_a$ constraints are strongly correlated, and the $\Lambda$CDM point, $(w_0,w_a)=(-1,0)$, remains well within the allowed region for both relations. At the same time, the corresponding $H_0$ and $\Omega_m$ posteriors remain consistent with their flat-$\Lambda$CDM counterparts. The present samples therefore provide only weak constraints on dynamical dark energy, but the extended model is useful for checking whether the behaviour found in flat $\Lambda$CDM persists when the expansion history is allowed greater freedom.

\begin{figure*}[htbp]
	\centering
	\includegraphics[width=0.85\textwidth]{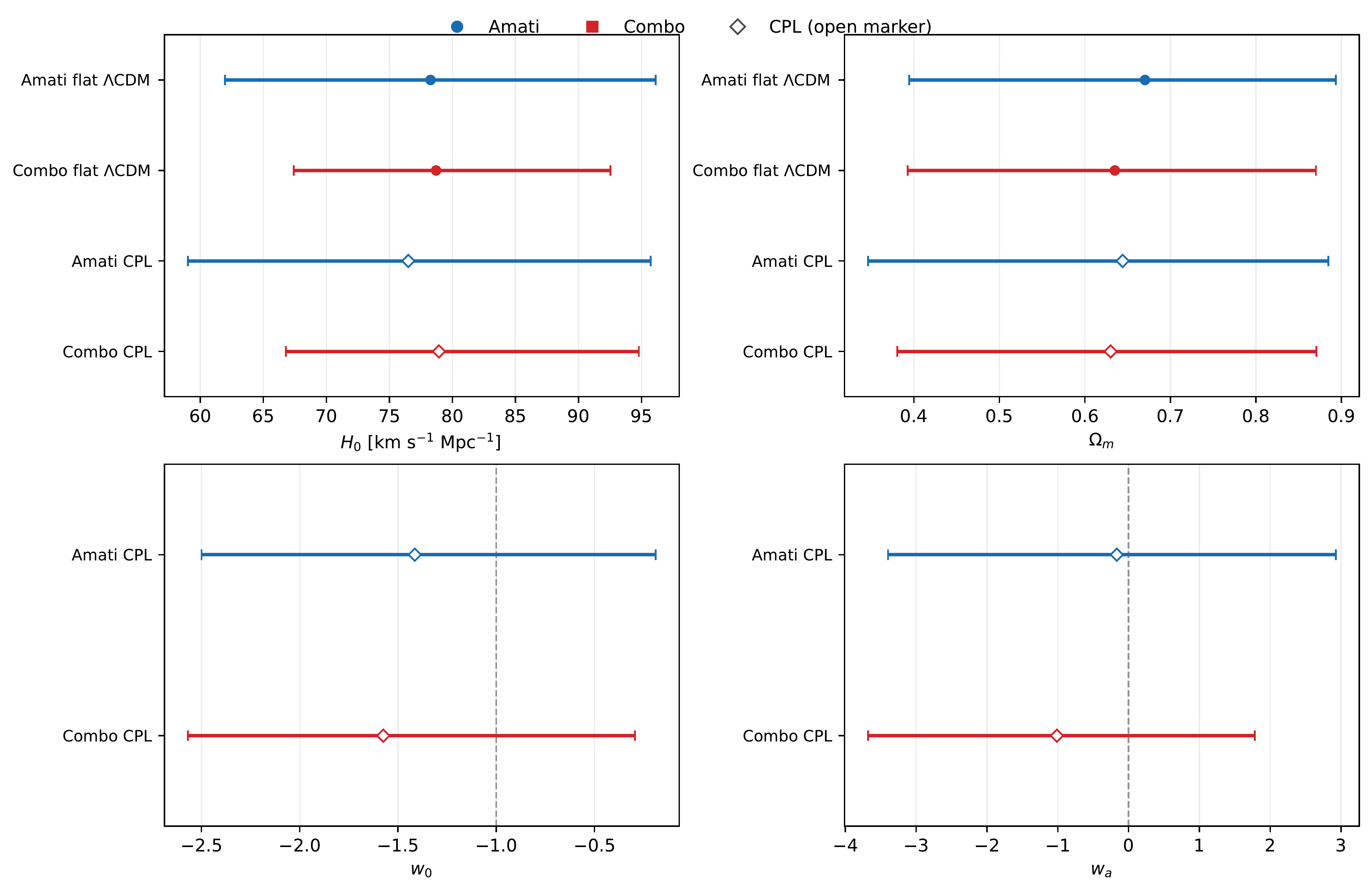}
	\caption{Reference cosmological constraints from the Amati and Combo relations. The upper panels show the constraints on $H_0$ and $\Omega_m$, while the lower panels show the $w_0w_a$CDM constraints on $w_0$ and $w_a$. Thick error bars indicate the central $68\%$ credible intervals. The two GRB analyses are shown separately but are not statistically independent.}
	\label{fig:baseline_constraints}
\end{figure*}

The calibrated GRB Hubble diagrams are shown in Figs.~\ref{fig:hubble_amati} and \ref{fig:hubble_combo}. The baseline Amati sample contains 19 calibration GRBs and 84 cosmological GRBs, reaching $z\approx8.26$, while the Combo sample contains 32 and 205 objects, respectively, and extends to $z\approx9.40$. The residuals remain broad compared with those of SNe Ia, as expected from the intrinsic dispersion of the GRB relations. Table~\ref{tab:cosmo_results} also shows that Combo has the smaller high-redshift scatter. The fitted intercepts and slopes change little between flat $\Lambda$CDM and $w_0w_a$CDM, so the differences discussed below are not associated with a substantial change in the fitted GRB correlations themselves.

\subsection{Redshift dependence and Combo tests}
\label{subsec:lambda_results}

\begin{figure}[htbp]
	\centering
	\includegraphics[width=0.5\textwidth]{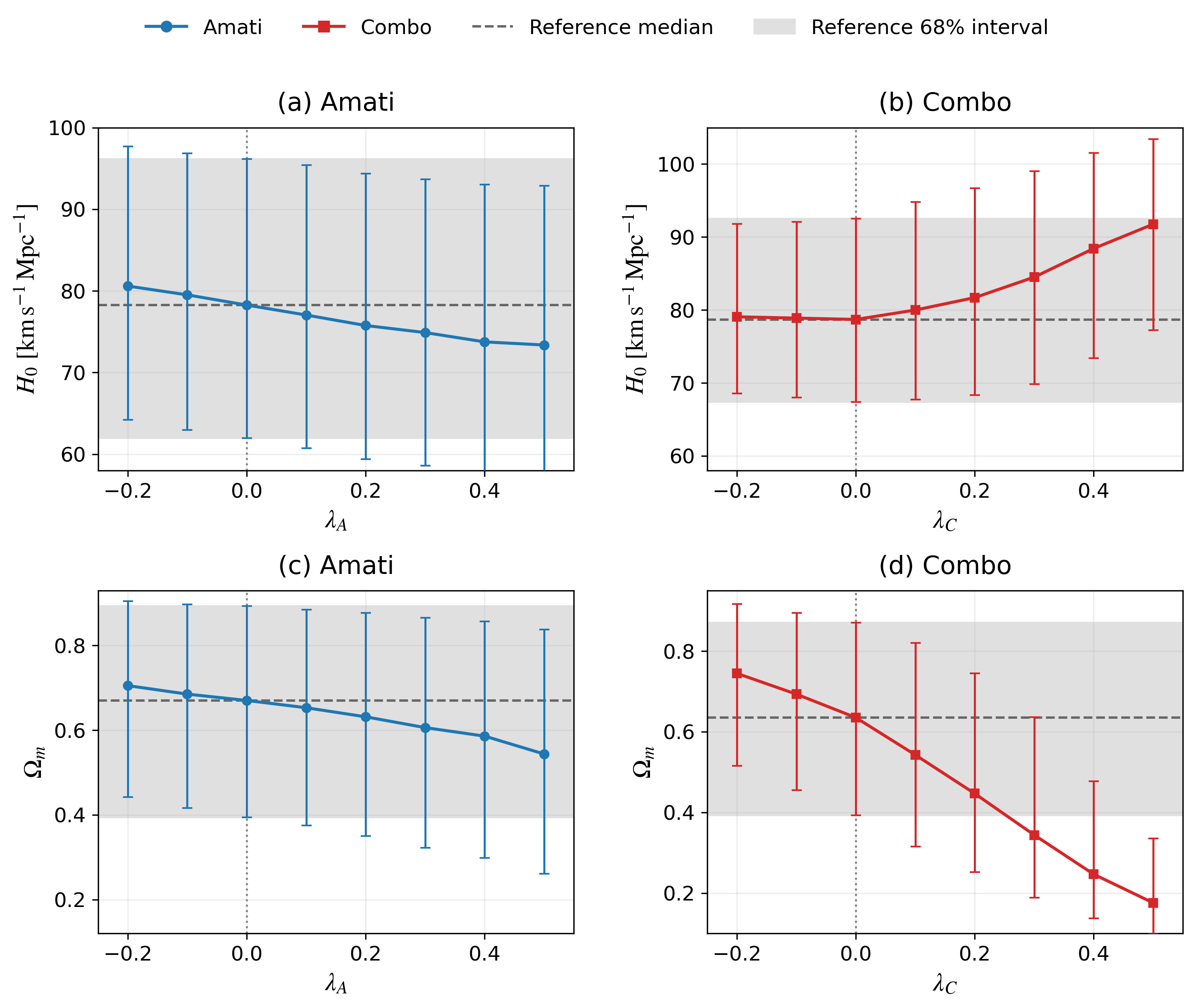}
	\caption{Dependence of the flat-$\Lambda$CDM constraints on the phenomenological redshift terms. The Amati constraints vary comparatively slowly, whereas the Combo $\Omega_m$ median decreases with increasing $\lambda_C$. The corresponding increase in the Combo $H_0$ median remains within the $1\sigma$ uncertainty of the reference constraint.}
	\label{fig:layer2_sensitivity_flatlcdm}
\end{figure}

We next examine the phenomenological redshift terms introduced in Sect.~\ref{subsec:robustness_tests}. The main flat-$\Lambda$CDM results are summarized in Table~\ref{tab:evolution_results} and Fig.~\ref{fig:layer2_sensitivity_flatlcdm}.

\begin{table}[htbp]
	\centering
	\caption{Flat-$\Lambda$CDM evolution tests.}
	\label{tab:evolution_results}
	\begin{tabular}{lccc}
		\toprule
		Analysis &$H_0$ &$\Omega_m$ &$\lambda_R$ \\
		\midrule
		Amati, reference &$78.26^{+17.87}_{-16.30}$ &$0.670^{+0.223}_{-0.276}$ &$0$ \\
		Amati, free $\lambda_A$ &$77.35^{+18.56}_{-17.04}$ &$0.620^{+0.255}_{-0.300}$ &$0.252^{+0.331}_{-0.371}$ \\
		Combo, reference &$78.71^{+13.83}_{-11.29}$ &$0.635^{+0.235}_{-0.242}$ &$0$ \\
		Combo, free $\lambda_C$ &$80.57^{+13.34}_{-11.29}$  &$0.679^{+0.218}_{-0.285}$ &$-0.0827^{+0.213}_{-0.199}$ \\
		\bottomrule
	\end{tabular}
\end{table}

For Amati, the posterior of the redshift parameter $\lambda_A$ is broad and includes zero, and the cosmological constraints are consistent with those obtained for $\lambda_A=0$. In the fixed-$\lambda_A$ scan, both $H_0$ and $\Omega_m$ decrease gradually as $\lambda_A$ increases, but the changes remain small compared with their posterior uncertainties. The dependence is stronger for Combo. The free-$\lambda_C$ posterior has a central $95\%$ interval of approximately $[-0.471,0.374]$ and does not exclude zero at this credible level. Figure~\ref{fig:layer2_sensitivity_flatlcdm} shows a steady decrease in the Combo $\Omega_m$ median with increasing $\lambda_C$. The corresponding $H_0$ median changes less at first and rises at the larger values of $\lambda_C$ considered in the scan. The relatively strong dependence of $\Omega_m$ on $\lambda_C$ indicates that the phenomenological redshift term is degenerate with the cosmological fit to the high-redshift Combo data. This dependence is less pronounced for Amati.

\begin{figure}[htbp]
	\centering
	\includegraphics[width=0.5\textwidth]{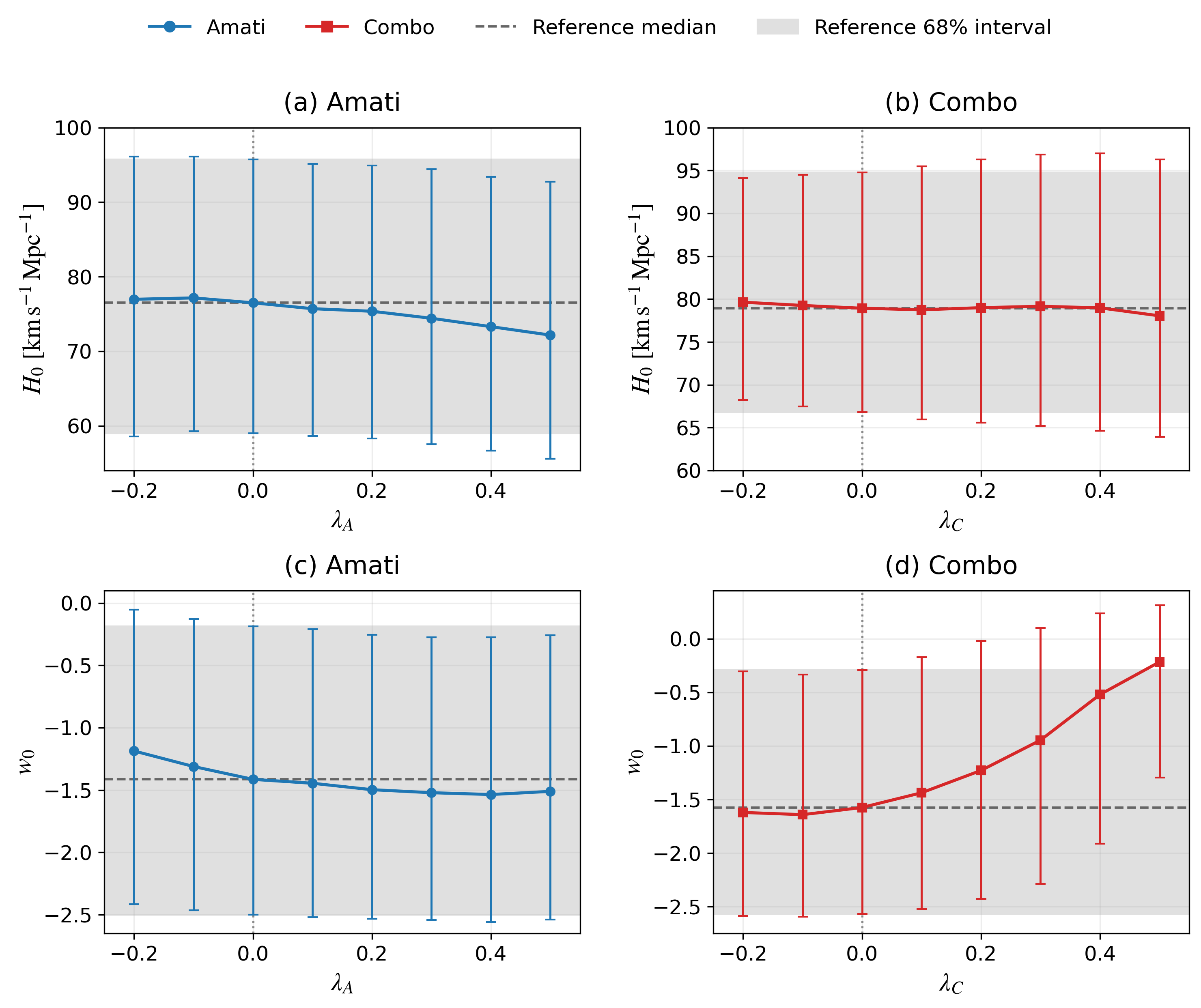}
	\caption{Same as Fig.~\ref{fig:layer2_sensitivity_flatlcdm}, but for $w_0w_a$CDM.The broader cosmological posteriors make the trends less pronounced. Amati remains comparatively stable, while Combo retains a tiny dependence on $\lambda_C$, particularly in the dark-energy parameters.}
	\label{fig:layer2_sensitivity_cpl}
\end{figure}

The corresponding $w_0w_a$CDM tests are shown in Fig.~\ref{fig:layer2_sensitivity_cpl}. Once $w_0$ and $w_a$ are allowed to vary, the broader posteriors weaken the association between the phenomenological redshift term and any single cosmological parameter. Amati remains comparatively stable, whereas Combo retains a moderate dependence on $\lambda_C$, most clearly in the dark energy parameters. The shifts remain small compared with the present posterior widths, so this behaviour should not be interpreted as evidence for evolving dark energy. It instead shows that the stronger redshift sensitivity of Combo persists when the assumed expansion history is enlarged beyond flat $\Lambda$CDM.

The Combo dependence could in principle arise if the much larger high-redshift sample substantially updates the correlation parameters constrained by the low-redshift calibration. We test this by transferring the calibration-only information to a high-redshift-only cosmological fit. For $\lambda_C=0$, the Gaussian-transfer and tightly constrained calibration cases give $\Omega_m=0.806$ and $0.830$, respectively. The high-$\Omega_m$ solution therefore persists even when the high-redshift sample is prevented from substantially changing the low-redshift relation. When $\lambda_C$ is allowed to vary, the corresponding values decrease to $\Omega_m=0.668$ and $0.667$, again consistent with the full free-$\lambda_C$ fit. The observed sensitivity is consequently associated mainly with the redshift dependence of the high-redshift Combo Hubble diagram rather than with reverse updating of the calibration relation.

To test this interpretation further, we construct semi-synthetic high-redshift Combo catalogues according to
\begin{equation}
	\log_{10}F_{0,\rm mock}=\log_{10}F_{0,\rm obs}+\left(\lambda_{\rm inj}-\lambda_{\rm rem}\right)\log_{10}\left[\frac{1+z}{1+z_{\rm ref}}\right],
	\label{eq:combo_injection}
\end{equation}
where $\lambda_{\rm rem}$ is the trend inferred from the observed free-$\lambda_C$ fit. We use $\lambda_{\rm inj}=0$, $0.32$, and $0.60$, leaving the calibration sample and the remaining measured GRB quantities unchanged. Since this test was constructed after the $\lambda_C$ dependence had been identified in the real data, it is intended to test the inferred degeneracy rather than to provide independent evidence for GRB evolution.

\begin{table}[htbp]
	\centering
	\caption{Combo injection--recovery results. The recovered $\lambda_C$ values are obtained from the free-$\lambda_C$ fits, and the remaining columns compare cosmological fits with $\lambda_C=0$ and with $\lambda_C$ sampled simultaneously.}
	\label{tab:combo_injection}
	\begin{tabular}{cccccc}
		\toprule
		$\lambda_{\rm inj}$ &$\lambda_C^{\rm rec}$ &$H_0^{\lambda_C=0}$ &$\Omega_m^{\lambda_C=0}$ &$H_0^{\lambda_C\ {\rm free}}$ &$\Omega_m^{\lambda_C\ {\rm free}}$ \\
		\midrule
		$0.00$ &$-0.002^{+0.219}_{-0.201}$ &$78.75$ &$0.682$ &$80.13$ &$0.666$ \\
		$0.32$ &$0.317^{+0.218}_{-0.199}$ &$81.14$ &$0.805$ &$79.80$ &$0.678$ \\
		$0.60$ &$0.587^{+0.200}_{-0.197}$ &$84.84$ &$0.857$ &$79.92$ &$0.680$ \\
		\bottomrule
	\end{tabular}
\end{table}

The injected values are recovered with a maximum median offset of about $0.013$. If the mock catalogues are instead analysed with $\lambda_C=0$, an increasingly positive injected trend shifts both $H_0$ and $\Omega_m$ upward, with the stronger response occurring in $\Omega_m$. Allowing $\lambda_C$ to vary absorbs most of this dependence. The experiment therefore reproduces the direction of the cosmological shift seen in the real Combo sample. Since the injected and fitted trends have the same functional form, however, it does not determine whether the observed redshift dependence arises from intrinsic GRB evolution, detector selection, follow-up incompleteness, or another population effect.

\subsection{Robustness and sample properties}
\label{subsec:results_robustness}

We finally consider analysis choices that do not change the functional form of the GRB relations. The calibration-boundary scan shows the largest shifts at low $z_{\rm cut}$, where fewer GRBs are available for calibration. For the Combo analysis, the posterior medians change from $(H_0,\Omega_m)=(65.08,0.786)$ at $z_{\rm cut}=0.55$ to $(79.07,0.622)$ at $z_{\rm cut}=0.745$, while the free-$\lambda_C$ analysis changes from approximately $(65.10,0.741)$ to $(78.71,0.635)$. For $z_{\rm cut}\gtrsim0.65$, the differences are small compared with the statistical uncertainties.

Replacing the separate calibration and cosmological scatter parameters with a common intrinsic scatter changes the historical Combo result only slightly, from $(H_0,\Omega_m)=(79.19,0.634)$ to approximately $(79.11,0.633)$. Wider flat-$\Lambda$CDM priors likewise produce only modest shifts: the free-$\lambda_C$ runs give approximately $(H_0,\Omega_m)=(83.57,0.379)$. Thus the high-$\Omega_m$ no-evolution result is not produced by the nominal prior boundaries or by the split-scatter treatment.

We also vary the Combo slope and modify the GRB samples through brightness and quality cuts, jackknife removal, and bootstrap resampling. None of these changes removes the stronger redshift sensitivity of Combo. To examine whether this behaviour may be related to the observed sample distribution, we further consider correlations between redshift and the flux- or fluence-related quantities, correlations with the corresponding luminosity or energy quantities, the Efron--Petrosian trend statistic, and empirical selection weights. These quantities are not included in the likelihood and are used only to characterize possible redshift-dependent structure in the observed samples.

\begin{figure*}[htbp]
	\centering
	\includegraphics[width=0.48\textwidth]{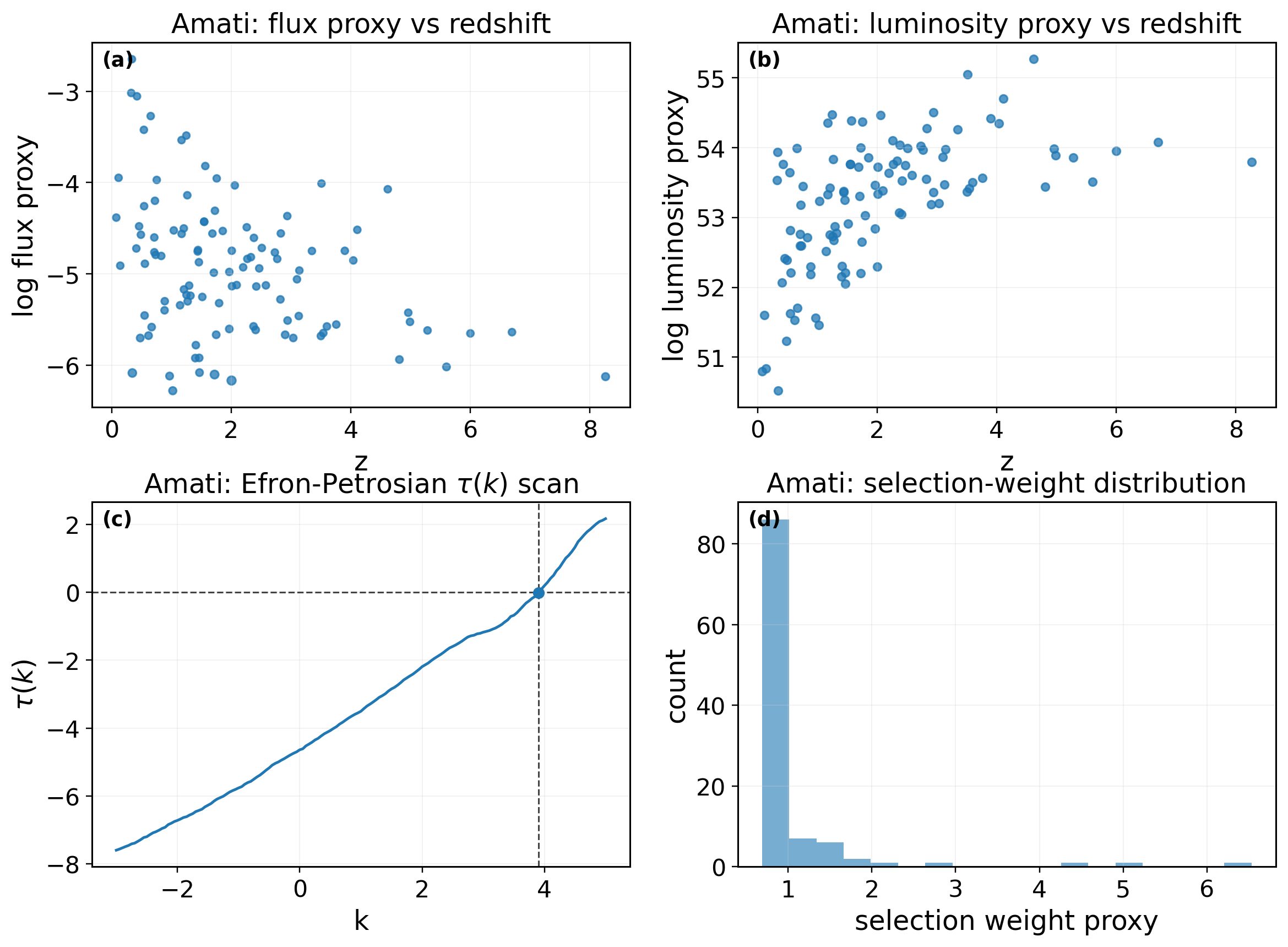}
	\hfill
	\includegraphics[width=0.48\textwidth]{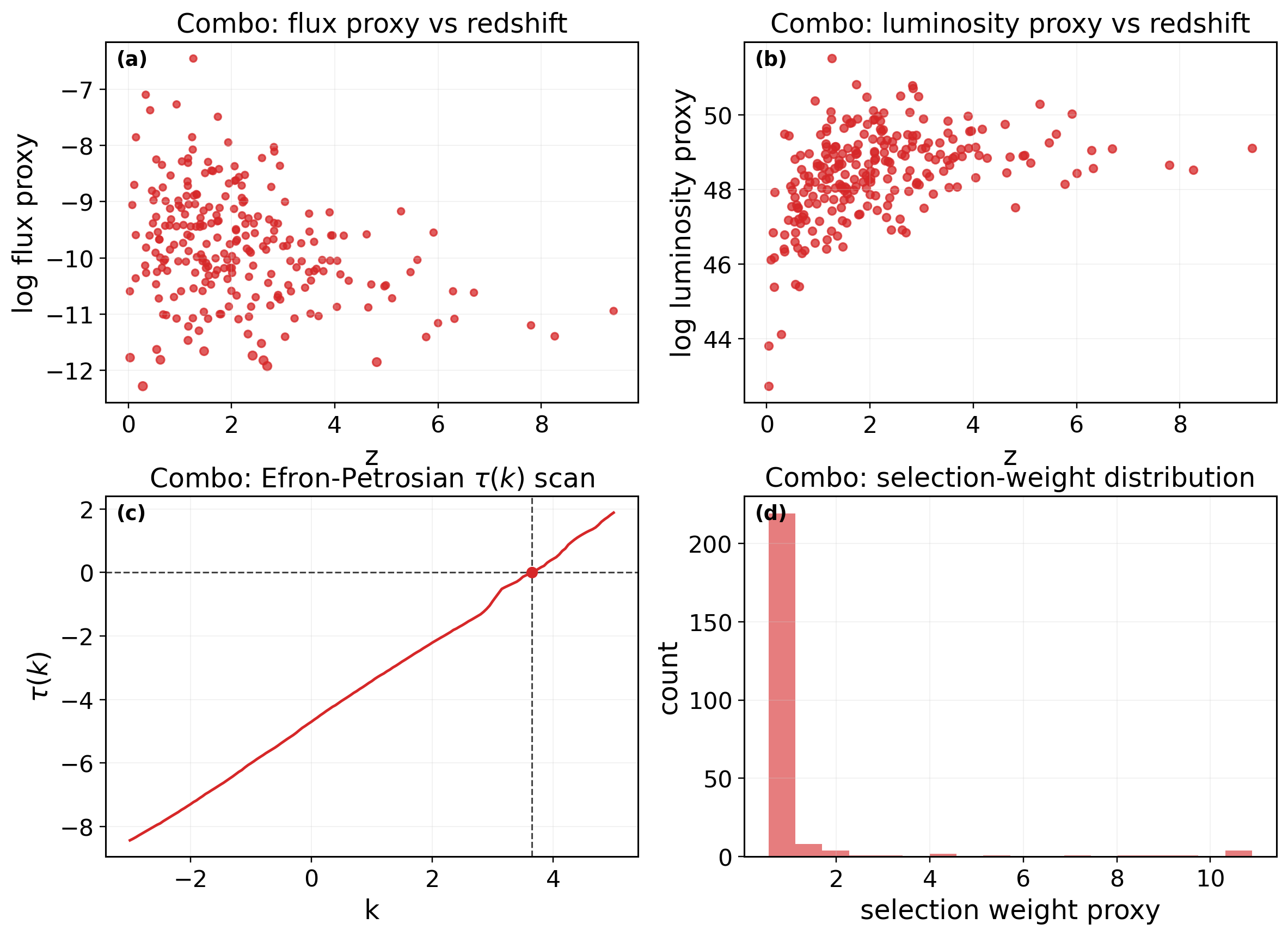}
	\caption{
		Selection diagnostics for the Amati (left) and Combo (right) samples.
		The panels show the observed redshift dependence of the flux- or fluence-related and luminosity- or energy-related quantities, the Efron--Petrosian trend statistics, and the empirical weight distributions.
		The Combo sample additionally requires a measurable X-ray plateau and is therefore affected by afterglow follow-up as well as prompt detection.
	}
	\label{fig:layer3_diagnostics}
\end{figure*}

The results are shown in Fig.~\ref{fig:layer3_diagnostics}. For Amati, the fluence proxy is anti-correlated with redshift, with a Spearman coefficient $\rho\approx-0.31$, whereas the energy-related quantity shows a positive correlation, $\rho\approx0.61$. The Efron--Petrosian scan gives $k\approx3.9$ for the approximate removal of the redshift--luminosity trend. These features are consistent with the expected behaviour of a flux-limited GRB sample. Combo shows similar but weaker rank correlations, with $\rho\approx-0.25$ for the flux--redshift relation and $\rho\approx0.47$ for the luminosity--redshift relation. Its empirical weight distribution is also broader, with median and maximum weights of approximately $0.55$ and $10.9$, compared with $0.72$ and $6.5$ for Amati. This is consistent with a less uniformly sampled redshift--luminosity distribution. In addition, the Combo sample requires both prompt-emission measurements and a measurable X-ray plateau, so its selection is affected by afterglow follow-up as well as prompt detection. These sample properties provide a plausible observational contribution to the stronger redshift dependence found for Combo, although the present diagnostics do not distinguish selection effects from intrinsic GRB evolution.

\section{Discussion}
\label{sec:discussion}

In this work, we have examined whether time-delay strong-lensing distances can provide an external calibration scale for GRB cosmology. The strong-lensing information is mapped onto a continuous distance--redshift relation with an ANN and subsequently used to calibrate the low-redshift Amati and Combo samples. The calibrated correlations then extend this distance scale to substantially higher redshifts. The inferred $H_0$ should therefore be interpreted as the propagation of the TDCOSMO absolute distance scale through the GRB ladder, rather than as an independent GRB-only determination of the Hubble constant. Similarly, although the ANN interpolation itself does not impose a parametric late-time expansion history, the calibration inherits the modelling assumptions and systematic uncertainties of the underlying strong-lensing analysis.

With the current eight-lens TDCOSMO sample, it is not possible to fully assess the impact of the assumed interpolation scheme. The ANN reconstruction is determined by the available strong-lensing distance measurements and the adopted network configuration, and different choices of interpolation methods may introduce additional systematic uncertainties into the reconstructed distance relation. Future strong-lensing surveys with a larger number of well-measured systems will allow a more detailed study of these effects and provide a more complete assessment of the uncertainty in the reconstructed distance scale.

The reference flat-$\Lambda$CDM constraints from Amati and Combo overlap within the present uncertainties, although the two samples partly overlap and should not be regarded as statistically independent. Their agreement nevertheless provides a useful comparison because the two relations involve different combinations of prompt and afterglow observables. Both analyses favour relatively high central values of $\Omega_m$, albeit with broad uncertainties. One possible point of interest is that the strong-lensing calibrators used here lie at substantially higher redshifts than the very local anchors commonly employed in traditional distance ladders. At these redshifts the expansion history has a greater influence on the distance relation, raising the possibility that calibration with intermediate-redshift lens systems may lead to a different parameter sensitivity from calibration with very low-redshift anchors. The present data are insufficient to establish such an effect, particularly because the Combo results also show a clear dependence on redshift-dependent GRB properties. A direct comparison between otherwise identical GRB analyses calibrated with low- and intermediate-redshift anchors would therefore be useful for determining whether part of the relatively high inferred $\Omega_m$ is associated with the location of the calibration sample.

Allowing the expansion history to vary within $w_0w_a$CDM substantially broadens the cosmological posteriors, while the $\Lambda$CDM point, $(w_0,w_a)=(-1,0)$, remains well within the allowed region for both GRB relations. The corresponding $H_0$ and $\Omega_m$ constraints remain consistent with the flat-$\Lambda$CDM results. The $w_0w_a$CDM analysis therefore provides a test of the dependence on the assumed late-time expansion history, but the present GRB data do not provide a meaningful measurement of dynamical dark energy. More generally, the GRB constraints remain much broader than those obtained from CMB, BAO, and SNe Ia. Their present value lies mainly in extending an externally calibrated distance relation to redshifts well beyond the dense SNe Ia regime rather than in providing competitive precision constraints.

The Amati and Combo relations show different behaviour under the redshift-dependent tests. Amati relies only on prompt-emission quantities and has a larger effective scatter, but its cosmological constraints change relatively little when $\lambda_A$ is varied. Combo has a smaller scatter and hence greater statistical weight, while its additional dependence on the X-ray plateau introduces sensitivity to afterglow detectability and follow-up completeness. The free-$\lambda_C$ analysis remains consistent with zero, so the data do not provide significant evidence for a physical evolution term. Nevertheless, the strong response of $\Omega_m$ to fixed values of $\lambda_C$ shows that redshift-dependent structure in the present Combo sample can be partially degenerate with cosmological parameters.

The calibration-transfer and injection--recovery tests help to locate the origin of this sensitivity. Restricting the Combo correlation parameters using the calibration-only posterior does not remove the high-$\Omega_m$ solution when $\lambda_C=0$ is imposed, indicating that the effect is not primarily caused by the high-redshift sample shifting the low-redshift calibration relation. In the semi-synthetic catalogues, an imposed positive redshift trend that is omitted from the fit produces an upward shift in the inferred cosmological parameters, particularly $\Omega_m$, while fitting $\lambda_C$ absorbs most of this shift. Since the injection test was constructed after the behaviour was identified in the real data and uses the same functional form as the fitted trend, it should be regarded as a test of this degeneracy rather than independent evidence for GRB evolution. It also cannot distinguish intrinsic evolution from detector selection, follow-up incompleteness, or other population effects.

The remaining tests give a consistent picture. The results become substantially more stable for $z_{\rm cut}\gtrsim0.65$, whereas lower calibration boundaries leave fewer GRBs to determine the low-redshift relation and produce larger shifts. Replacing the separate calibration and cosmological scatter parameters with a common scatter or widening the cosmological priors produces only modest changes. Variations of the Combo slope, sample cuts, jackknife tests, and bootstrap resampling likewise do not remove its stronger redshift sensitivity. The rank-correlation, Efron--Petrosian, and empirical-weight analyses further indicate non-uniform redshift-dependent sampling, particularly for Combo. These results support the interpretation that the present Combo constraints are more sensitive to sample properties than the Amati constraints, without identifying a unique physical origin for the effect.

The strong-lensing calibration itself remains subject to modelling uncertainties. Time-delay cosmography depends on the lens mass distribution, stellar kinematics, internal and external mass-sheet transformations, and line-of-sight structure. These effects are incorporated in the TDCOSMO analysis from which the distances used here are derived, but any residual uncertainty is inherited by the GRB calibration. The ANN does not remove these dependencies; it transfers the lensing distance information to the GRB calibration redshifts. In the baseline analysis, calibration is restricted to $0.2283\leq z\leq0.745$, so the nominal cosmological constraints do not rely on unsupported ANN extrapolation. Since the reconstructed distances at different GRB redshifts originate from the same small set of lens systems, their uncertainties are strongly correlated, making the full ANN covariance an essential part of the calibration likelihood.

The dominant astrophysical limitation is the GRB population itself. Although separate intrinsic-scatter parameters are allowed for the calibration and cosmological subsets, the present likelihood does not contain a complete population model for detector thresholds, luminosity evolution, redshift completeness, or X-ray plateau recovery. Prompt-emission selection affects the Amati sample, while Combo additionally requires sufficiently complete X-ray follow-up to identify and characterize the plateau. The phenomenological $\lambda_R$ parameters quantify how such redshift-dependent effects can propagate into the cosmological inference, but they do not determine their origin. A more complete treatment will require the GRB luminosity function and observational selection function to be incorporated directly into the hierarchical likelihood.

These limitations place our results in the broader context of GRB cosmology. Analyses using different calibration strategies generally find $\Lambda$CDM to remain viable once calibration uncertainties are propagated consistently \citep{2020A&A...641A.174L}, whereas studies based on DESI, OHD, or BAO calibration have reported mild preferences for dynamical dark energy in particular data combinations \citep{2024JCAP...12..055A,2025JHEAp..4600348A}. More generally, recent cosmological analyses have shown that apparent evidence for evolving dark energy can depend on the adopted data combination and calibration treatment \citep{2025PhRvD.111b3512C,2024A&A...690A..40L}. The present results show no statistically meaningful departure from $\Lambda$CDM and emphasize that calibration and redshift-dependent systematics must be understood before broad high-redshift constraints are interpreted in terms of new cosmological physics.

Future strong-lensing samples should make this comparison substantially more informative. Rubin/LSST, Euclid, and the Nancy Grace Roman Space Telescope are expected to increase the number of well-characterized time-delay systems and extend their redshift coverage \citep{2010MNRAS.405.2579O,2020RNAAS...4..190W,2022A&ARv..30....8T,2023MNRAS.524.5446T,2025A&A...697A...1E,2025ApJ...986...42W}. A denser set of lens anchors would improve the distance reconstruction and allow the calibration scale to be compared across lens redshift and population. Together with better characterized GRB selection functions, this would make it possible to test whether the relatively high matter-density values found here are associated with the calibration scale, the redshift distribution of the GRBs, or a combination of the two. In this sense, Amati provides a comparatively stable but noisy high-redshift distance indicator, while Combo offers smaller statistical scatter at the cost of greater sensitivity to redshift-dependent sample properties.

\section{Conclusions}
\label{sec:conclusions}

We have investigated whether time-delay strong-lensing distances can provide an external calibration scale for GRB cosmology. Using the TDCOSMO2025 strong-lensing information, we reconstruct the low-redshift distance--redshift relation with an ANN and calibrate the Amati and Combo correlations within the redshift interval directly covered by the lensing data. The calibrated relations are then applied to high-redshift GRBs in a hierarchical Bayesian analysis. Our main conclusions are as follows.

(i) Strongly lensed quasars provide an external calibration route for GRB cosmology that is independent of assuming a specific background cosmology for the GRB calibration itself. The correlated ANN reconstruction uncertainty is propagated into the calibration likelihood, and the baseline analysis is restricted to $0.2283\leq z\leq0.745$ to avoid unsupported ANN extrapolation. The resulting Amati and Combo Hubble diagrams extend the calibrated distance scale to $z\gtrsim8$ and $z\gtrsim9$, respectively.

(ii) In flat $\Lambda$CDM, the reference Amati analysis gives $H_0=78.26^{+17.87}_{-16.30}\,\mathrm{km\,s^{-1}\,Mpc^{-1}}$ and $\Omega_m=0.670^{+0.223}_{-0.276}$, while the Combo gives $H_0=78.71^{+13.83}_{-11.29}\,\mathrm{km\,s^{-1}\,Mpc^{-1}}$ and $\Omega_m=0.635^{+0.235}_{-0.242}$. The two relations give consistent constraints within the present uncertainties, although their samples partly overlap and should not be regarded as statistically independent. The inferred $H_0$ is also not an independent GRB-only determination, since its absolute scale is inherited from TDCOSMO and propagated through the ANN and GRB likelihoods.

(iii) Extending the analysis to $w_0w_a$CDM substantially broadens the cosmological posteriors. The $\Lambda$CDM point, $(w_0,w_a)=(-1,0)$, remains well within the allowed region for both relations, and the corresponding $H_0$ and $\Omega_m$ constraints remain consistent with the flat-$\Lambda$CDM results. The present GRB data therefore provide no significant evidence for dynamical dark energy.

(iv) Amati and Combo show different sensitivities to redshift-dependent effects. The Amati constraints remain relatively stable when the phenomenological redshift term is varied, although the relation has a larger intrinsic scatter. Combo has a smaller effective scatter, but its inferred cosmological parameters are more sensitive to the treatment of redshift dependence. In particular, fixing $\lambda_C=0$ shifts $\Omega_m$ toward higher values, while allowing $\lambda_C$ to vary substantially reduces this shift. The free-$\lambda_C$ posterior remains broad and only shift the zero within $2\sigma$, so the present data do not provide significant evidence for physical evolution of the Combo relation.

(v) The calibration-transfer and injection--recovery tests show that the high-$\Omega_m$ no-evolution solution is not primarily caused by the high-redshift sample shifting the low-redshift Combo calibration. The semi-synthetic tests further show that an unmodelled positive redshift trend can shift the inferred cosmological parameters upward, particularly $\Omega_m$, while allowing $\lambda_C$ to vary removes most of this dependence. Together with the calibration-boundary, scatter, prior, resampling, and selection tests, these results indicate that the present Combo constraints are more sensitive than the Amati constraints to redshift-dependent sample properties. The analysis does not distinguish intrinsic GRB evolution from detector selection, follow-up incompleteness, or other population effects.

The present precision is limited by the number and redshift distribution of the strong-lensing anchors, residual lens-modelling uncertainties, the correlated ANN reconstruction, and the intrinsic scatter and selection properties of the GRB samples. The relatively high central values of $\Omega_m$ also motivate future comparisons between GRB calibrations based on very low-redshift and intermediate-redshift distance anchors, which may help separate calibration-scale effects from redshift-dependent GRB systematics.

Future progress will therefore require improvements on both sides of the calibration. Rubin/LSST, Euclid, and the Nancy Grace Roman Space Telescope should substantially enlarge the sample and redshift coverage of well-characterized time-delay lenses. On the GRB side, SVOM \citep{2011CRPhy..12..298P,2016arXiv161006892W,2022IJMPD..3130008A} and THESEUS \citep{2018AdSpR..62..191A,2018AdSpR..62..662S,2021arXiv210208702A,2021ExA....52..183A} are expected to provide larger and more homogeneous samples with improved prompt and afterglow measurements. Combined with explicit population-level modelling of detector thresholds and follow-up completeness, these developments will determine how far strong-lensing-calibrated GRBs can improve measurements of the expansion history beyond the current SNe Ia redshift range.

%\clearpage

%%%%%%%%%%%%%%%%%%%%%%%%%%%%%%%%%%%%%%%%%%%%%%%%%%%%%%%%%%%%%%

%\newpage 

%%%%%%%%%%%%%%%%%%%%%%%%%%%%%%%%%%%%%%%%%%%%%%%%%%%%%%%%%%%%%%
\begin{acknowledgements}
Wei Hong is supported by China Scholarship Council (File No.202506040057). Luca Izzo acknowledges financial support from the INAF Data Grant Program 'YES' (PI: Izzo) {\it Multi-wavelength and multi messenger analysis of relativistic supernovae}. Tong-Jie Zhang is supported by National SKA Program of China, No. 2022SKA0110202, and the China Manned Space Program with grant No. CMS-CSST-2025-A01 1 and China Scholarship Council (File No. 202606040159). 
\end{acknowledgements}

\bibliographystyle{aa}
\bibliography{GRBCC}

\begin{appendix}
	
% =========================================================
\section{Strong-lensing distance preprocessing}
\label{app:tdcosmo_preprocessing}
% =========================================================
	
This Appendix summarizes the construction of the strong-lensing distance catalogue used as input to the ANN reconstruction. The preprocessing propagates the TDCOSMO2025 time-delay and stellar-kinematic information into deflector angular-diameter distances and subsequently converts them into luminosity distances and distance moduli at the lens redshifts.
	
For each lens system, the joint time-delay and kinematic information can be written in the form
\begin{equation}
D_{\rm d}=D_{\Delta t}(1-\kappa_{\rm ext})\frac{c^2J(\xi_{\rm lens},\beta_{\rm ani},\lambda_{\rm int})}{(1+z_{\rm d})\sigma_{\rm los}^2}.
		\label{eq:app_Dd_from_Ddt}
	\end{equation}
The resulting $D_{\rm d}$ posterior therefore inherits the lens-model, line-of-sight, and stellar-kinematic uncertainties included in the TDCOSMO analysis. The angular-diameter distance is converted to luminosity distance through the distance-duality relation $D_L(z_{\rm d})=(1+z_{\rm d})^2D_{\rm d}$, and the corresponding distance modulus is
$\mu=5\log_{10}\left(\frac{D_L}{\rm Mpc}\right)+25$. The resulting lens catalogue used for the ANN reconstruction is listed in Table~\ref{tab:app_tdcosmo_ann_input}.
\begin{table}[htbp]
	\centering
	\scriptsize
\caption{Strong-lensing distance catalogue used as input to the ANN reconstruction. The quoted values are medians with central $68\%$ intervals. The listed values are independent strong-lensing distance measurements, and the monotonicity condition is applied to the reconstructed distance--redshift relation in the ANN analysis (Appendix~\ref{app:ann_physical}).}
	\label{tab:app_tdcosmo_ann_input}
	\begin{tabular}{lcccc}
	\hline\hline
	Lens & $z_{\rm d}$ & $D_{\rm d}$ [Mpc] & $D_L$ [Mpc] & $\mu$ \\
\hline
WGD2038$-$4008 & 0.2283 & $993.3\pm236.6$ & $1498.5\pm357.0$ & $40.878\pm0.517$ \\
RXJ1131$-$1231 & 0.2950 & $808.9\pm54.6$ & $1356.6\pm91.6$ & $40.662\pm0.147$ \\
PG1115+080 & 0.3110 & $986.8\pm153.9$ & $1696.1\pm264.5$ & $41.147\pm0.339$ \\
HE0435$-$1223 & 0.4546 & $1209.2\pm143.2$ & $2558.5\pm302.9$ & $42.040\pm0.257$ \\
DES0408$-$5354 & 0.5970 & $1659.4\pm186.6$ & $4232.1\pm475.8$ & $43.133\pm0.244$ \\
B1608+656 & 0.6304 & $858.7\pm88.3$ & $2282.7\pm234.8$ & $41.792\pm0.223$ \\
WFI2033$-$4723 & 0.6575 & $1693.0\pm228.0$ & $4651.2\pm626.4$ & $43.338\pm0.292$ \\
SDSS1206+4332 & 0.7450 & $1256.3\pm225.2$ & $3825.6\pm685.7$ & $42.914\pm0.389$ \\
\hline
	\end{tabular}
\end{table}

\begin{figure*}[htbp]
	\centering
	\includegraphics[width=0.8\textwidth]{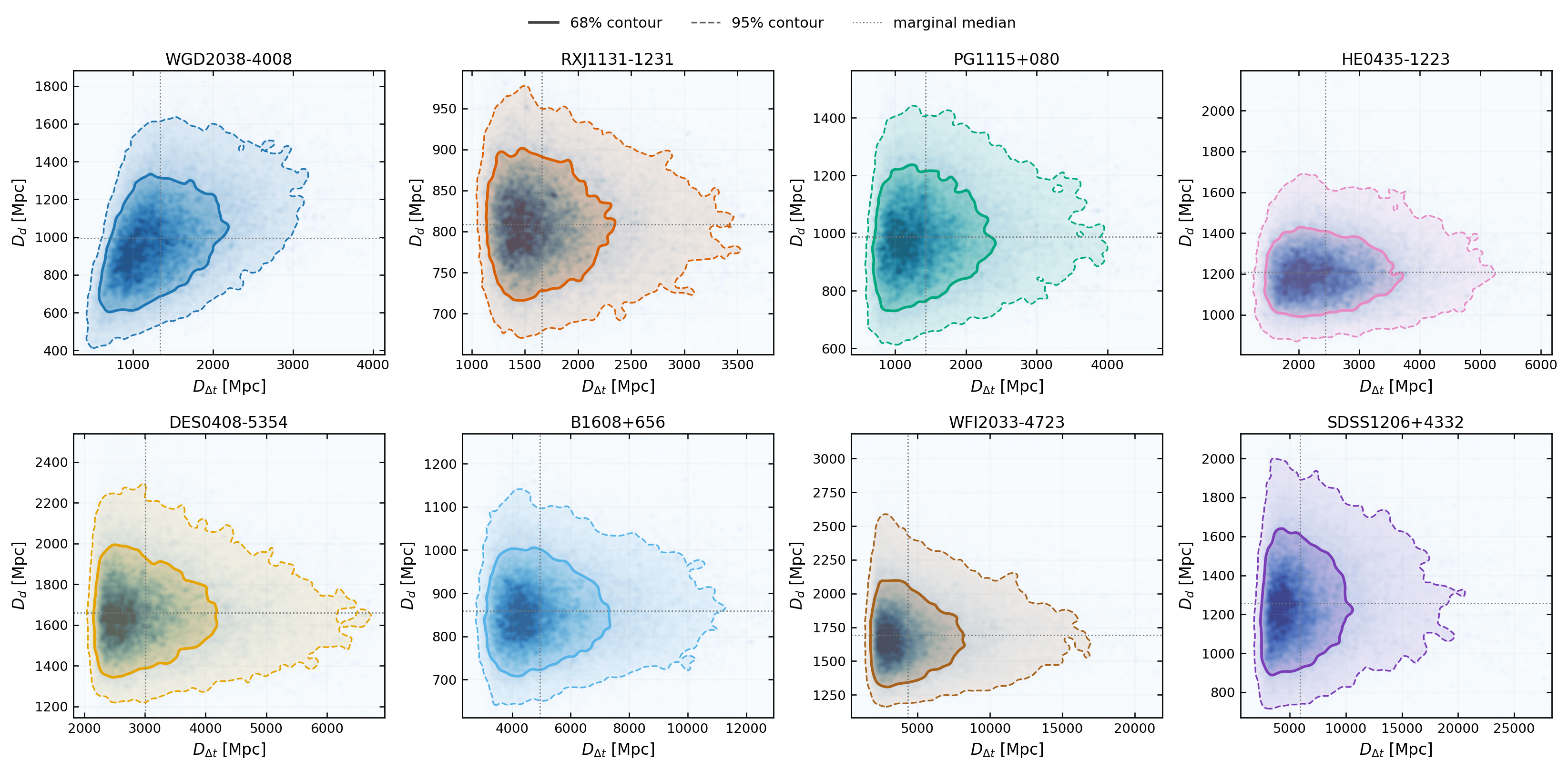}
	\caption{Joint posterior distributions in the $D_{\Delta t}$--$D_{\rm d}$ plane for the eight TDCOSMO2025 lens systems. Solid and dashed contours indicate the $68\%$ and $95\%$ credible regions, respectively, while the dotted lines mark the marginal posterior medians.}
		\label{fig:app_tdcosmo_dd_panel}
\end{figure*}

Figure~\ref{fig:app_tdcosmo_dd_panel} shows the joint
$D_{\Delta t}$--$D_{\rm d}$ posteriors for the eight systems. These distributions provide a direct check of the propagation from the time-delay and kinematic information to the distances subsequently supplied to the ANN. They are used to construct and validate the lensing-distance catalogue rather than as an additional cosmological likelihood in the GRB analysis.
	
% =========================================================
\section{ANN reconstruction and validation}
\label{app:ann_validation}
% =========================================================

The ANN is used to interpolate the strong-lensing distance information across the redshift range required for the GRB calibration. Because only eight lens systems are presently available, we impose physical requirements on the accepted reconstructions, examine the dependence on the network architecture and individual lens systems, and propagate the resulting reconstruction uncertainty through the full predictive covariance.

% =========================================================
\subsection{Reconstruction variables and physical requirements}
\label{app:ann_physical}
% =========================================================

The input catalogue is written as $\mathcal{D}_{\rm L}=\left\{z_i,D_{L,i},\mu_i,\sigma_{\mu,i}\right\}_{i=1}^{N_{\rm L}}$.
To reduce the dynamic range of the target quantity, we use $x_i=\ln(1+z_i)$ and $u_i=\ln\left[\frac{D_L(z_i)}{z_i(1+z_i)}\right]$. The network reconstructs $u(x)$, from which the luminosity distance is obtained as $D_L(z)=z(1+z)\exp[u(x)]$.

Every accepted reconstruction is required to satisfy $D_L(z)>0$ and $\frac{{\rm d}D_L}{{\rm d}z}>0$ throughout the lens-supported calibration interval. Reconstructions violating either condition are excluded from the final ensemble. The low-redshift boundary condition $D_L(0)=0$ is imposed analytically through the adopted parametrization and the behaviour $D_L(z)\propto z$ as $z\rightarrow0$. And a finite distance modulus is therefore not assigned at $z=0$. These requirements prevent unphysical ANN realizations from entering either the reconstructed mean distance relation or its covariance.

% =========================================================
\subsection{Network selection}
\label{app:ann_model_selection}
% =========================================================

Given the small number of lensing anchors, the ANN architecture is compared across several feed-forward configurations using complementary statistics rather than selected from a single numerical minimum.

For candidate model $M_k$, with residual vector $\mathbf{r}^{(k)}=\boldsymbol{\mu}_{\rm L}-\boldsymbol{\mu}_{\rm L}^{(k)}$, we calculate $D_{\chi^2}^{(k)}=\left(\mathbf{r}^{(k)}\right)^{\rm T}\mathbf{C}_{\rm L}^{-1}\mathbf{r}^{(k)}$, the mean prediction error $D_{\rm E}^{(k)}=\frac{1}{N_{\rm L}}\left[\sum_i\left(\mu_i-\mu_i^{(k)}\right)^2\right]^{1/2}$, and the Gaussian likelihood statistic $D_{\rm GL}^{(k)}=\frac{1}{2}\left[\left(\mathbf{r}^{(k)}\right)^{\rm T}\mathbf{C}_{\rm L}^{-1}\mathbf{r}^{(k)}+\ln|\mathbf{C}_{\rm L}|+N_{\rm L}\ln(2\pi)\right]$. The latter is used only for relative model comparison and is not interpreted as a Bayesian evidence.

Within the ABC rejection step, a realization is retained when $\rho(\mathbf{s}_{\rm sim},\mathbf{s}_{\rm obs})<\varepsilon$, where $\mathbf{s}$ denotes the summary statistics, $\rho$ the adopted distance measure, and $\varepsilon$ the tolerance. The tolerance is progressively reduced until the relative preference among the surviving network configurations becomes stable.

The comparison is supplemented by the risk statistic ${\rm Risk}^{(k)}=\sum_i\left[\mu_i-\mu_i^{(k)}\right]^2+\sum_i{\rm Var}\left[\mu_i^{(k)}\right]$, which accounts for both the mismatch with the lensing distances and the variability among repeated reconstructions. We also use leave-one-out tests to assess sensitivity to individual lens systems.

\begin{figure}[htbp]
	\centering
	\includegraphics[width=1\linewidth]{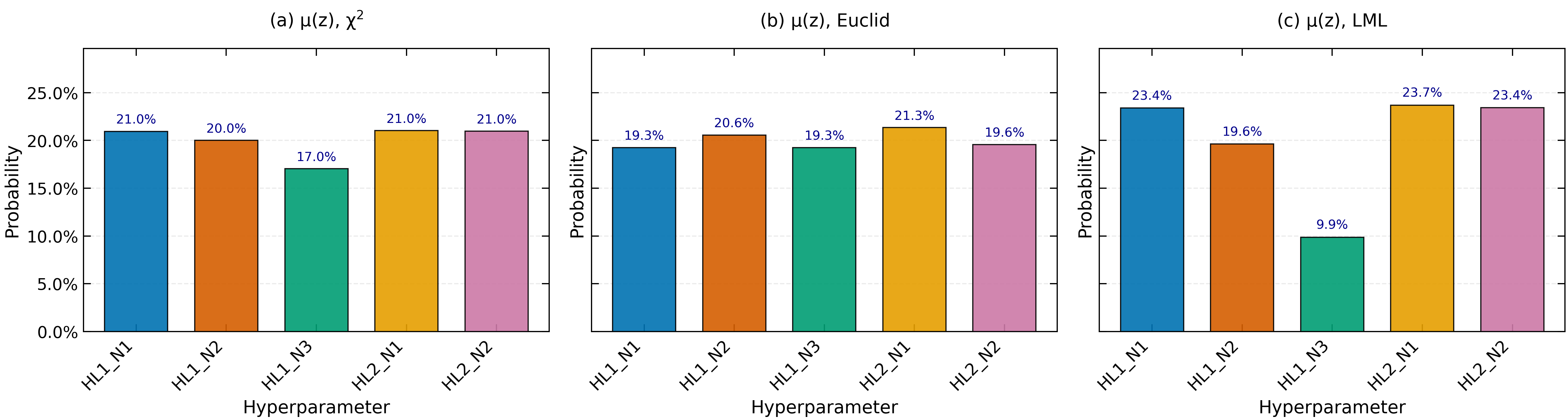}
	\caption{Relative ABC-rejection weights for the tested ANN hyperparameters using the $\chi^2$, Euclidean prediction-error, and Gaussian-likelihood statistics.}
	\label{fig:app_ann_probabilities}
\end{figure}

\begin{figure*}[htbp]
	\centering
	\includegraphics[width=0.85\textwidth]{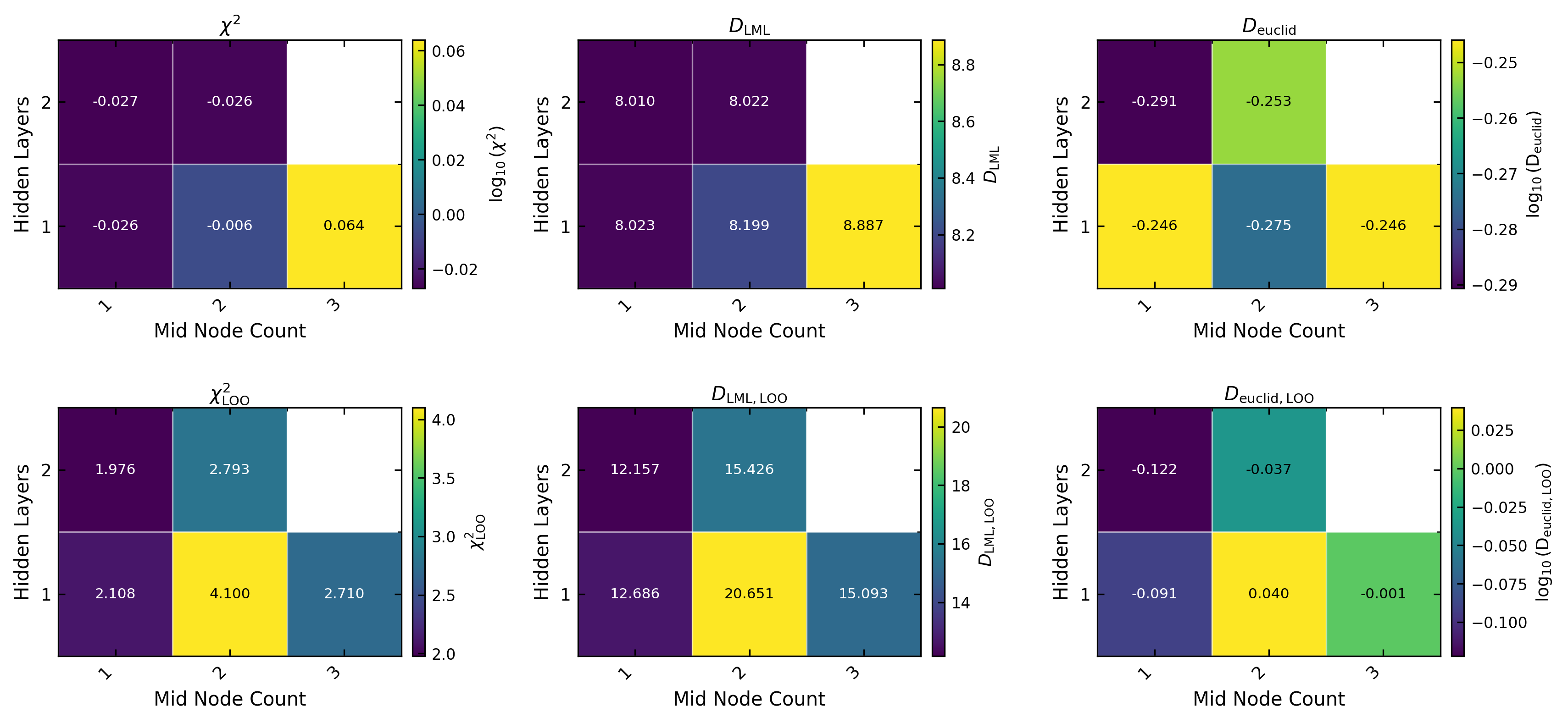}
	\caption{Hyperparameter diagnostics for the ANN reconstruction. The upper row shows the full-sample statistics and the lower row the corresponding leave-one-out quantities.}
	\label{fig:app_ann_heatmaps}
\end{figure*}

\begin{figure*}[htbp]
	\centering
	\includegraphics[width=0.85\linewidth]{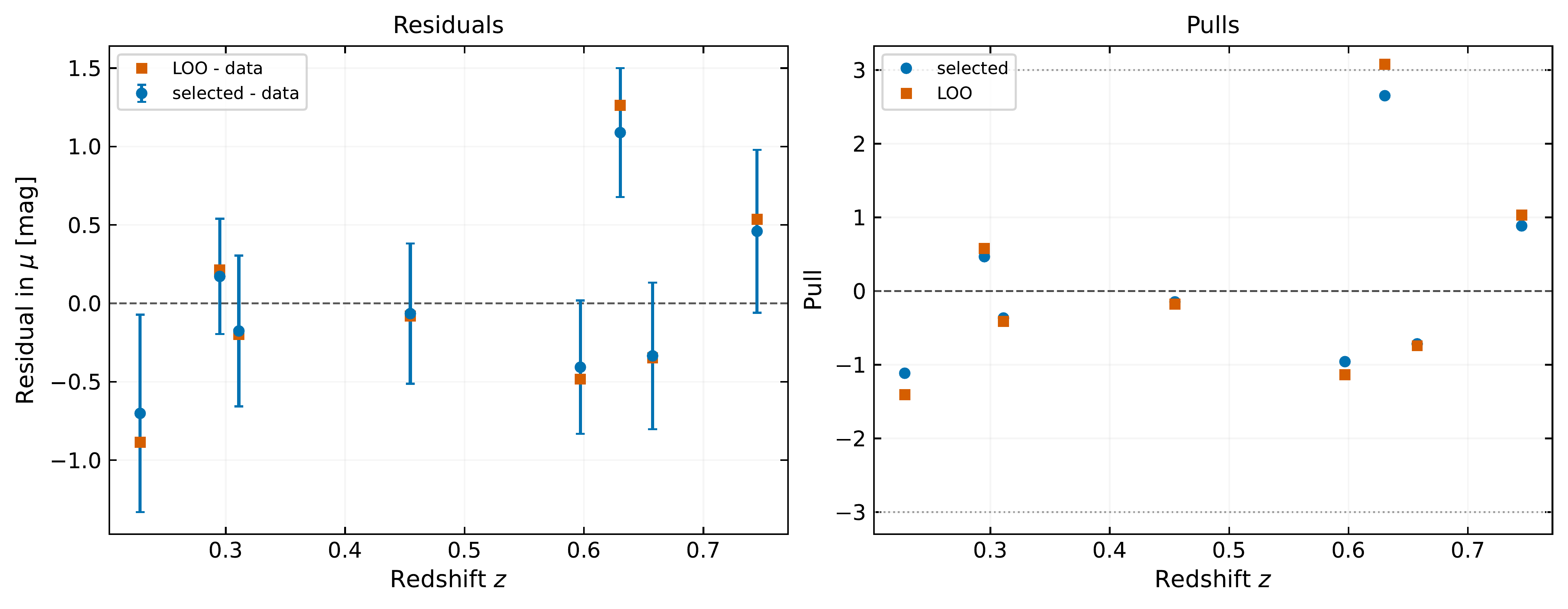}
	\caption{
		Residual and leave-one-out checks for the selected ANN reconstruction. No individual lens dominates the reconstructed relation, and the normalized residuals remain within the plotted uncertainty range.}
	\label{fig:app_ann_residuals}
\end{figure*}

Figures~\ref{fig:app_ann_probabilities}--\ref{fig:app_ann_residuals} summarize these diagnostics. The ABC weights are distributed over several architectures, the full-sample and leave-one-out statistics vary smoothly across the tested hyperparameter space, and no individual lens produces a dominant change in the selected reconstruction.

% =========================================================
\subsection{Covariance construction and numerical validation}
\label{app:ann_covariance}
% =========================================================

The accepted ANN ensemble is subsequently evaluated at the GRB calibration redshifts. Since all reconstructed distances are derived from the same underlying lens catalogue, their uncertainties are correlated and cannot be treated as independent measurements.

For an ensemble of $N_{\rm acc}$ accepted ANN realizations, the mean reconstructed distance modulus is $\bar{\mu}_{\rm ANN}(z_a)=\frac{1}{N_{\rm acc}}\sum_{r=1}^{N_{\rm acc}}\mu_r(z_a)$, and the predictive covariance is
\begin{equation}
	C_{\rm ANN}(z_a,z_b)
	=
	\frac{1}{N_{\rm acc}-1}
	\sum_{r=1}^{N_{\rm acc}}
	\left[
	\mu_r(z_a)-\bar{\mu}_{\rm ANN}(z_a)
	\right]
	\left[
	\mu_r(z_b)-\bar{\mu}_{\rm ANN}(z_b)
	\right].
	\label{eq:app_ann_covariance}
\end{equation}
The ensemble includes repeated realizations of both the strong-lensing input catalogue and the ANN reconstruction, thereby propagating the lensing-distance uncertainty and reconstruction variability.

The corresponding correlation coefficients are $\rho_{ab}=\frac{C_{\rm ANN}(z_a,z_b)}{\sqrt{C_{\rm ANN}(z_a,z_a)C_{\rm ANN}(z_b,z_b)}}$. The reconstructed distances are correlated because they are derived from the same set of strong-lensing measurements, and the full covariance matrix is retained in the calibration likelihood. The effective covariance is $\mathbf{C}_{\rm ANN,eff}=\mathbf{C}_{\rm ANN}+\sigma_{\rm rec}^{2}\mathbf{I}$, where $\sigma_{\rm rec}$ accounts for the residual reconstruction dispersion not included in the ensemble covariance. The residual reconstruction variance is estimated from the scatter not described by the ensemble covariance, giving $\sigma_{\rm rec}^{2}=0.112525$. For a given calibration sample, $ \left[\mathbf{C}_{\rm ANN,cal}\right]_{ij}=C_{\rm ANN,eff}\left(z_{{\rm GRB},i},z_{{\rm GRB},j}\right)$, with the GRB measurement errors and intrinsic scatter subsequently added as described in Sect.~\ref{subsec:hierarchical_likelihood}.

Because the reconstructed distances can be strongly correlated, we inspect the eigenspectrum $\mathbf{C}\mathbf{v}_k=\lambda_k\mathbf{v}_k$ and require all numerically resolved eigenvalues to be positive. We also monitor the condition number $\kappa(\mathbf{C})=\lambda_{\max}/\lambda_{\min}$ and the log-determinant $\ln|\mathbf{C}|=\sum_k\ln\lambda_k$, rather than using the raw determinant as the principal indicator of near-singularity.

When required for stable matrix factorization, a small numerical jitter is added $\mathbf{C}_{\rm num}=\mathbf{C}+\epsilon_{\rm jit}\mathbf{I}$ and $\epsilon_{\rm jit}=10^{-10}$. This regularization is negligible compared with the statistical variances. No covariance eigenmode is discarded, and the full matrix is retained rather than replaced by a diagonal approximation. The Gaussian likelihood is evaluated through Cholesky factorization rather than by explicitly computing $\mathbf{C}^{-1}$ or the determinant. Table~\ref{tab:ann_covariance_diagnostics} summarizes the numerical properties of the covariance matrices used in the production analysis.

\begin{table}[htbp]
	\centering
	\caption{Numerical diagnostics of the ANN covariance propagated to the GRB calibration samples. $\mathbf{C}_{\rm member}$ denotes the covariance of the interpolated ANN ensemble, while$\mathbf{C}_{\rm eff}=\mathbf{C}_{\rm member}+\sigma_{\rm rec}^2\mathbf{I}$.}
	\label{tab:ann_covariance_diagnostics}
	\begin{tabular}{lccccc}
		\toprule
		Sample &
		$N_{\rm cal}$ &
		${\rm Tr}(\mathbf{C}_{\rm member})$ &
		${\rm Tr}(\mathbf{C}_{\rm eff})$ &
		$\sigma_{\rm rec}^2$ &
		$\epsilon_{\rm jit}$ \\
	\midrule
	Amati & 19 & 0.45860 & 2.59657 & 0.112525 & $10^{-10}$ \\
	Combo & 32 & 0.77290 & 4.37369 & 0.112525 & $10^{-10}$ \\
		\bottomrule
	\end{tabular}
\end{table}

\begin{figure}[htbp]
	\centering
	\includegraphics[width=0.4\textwidth]{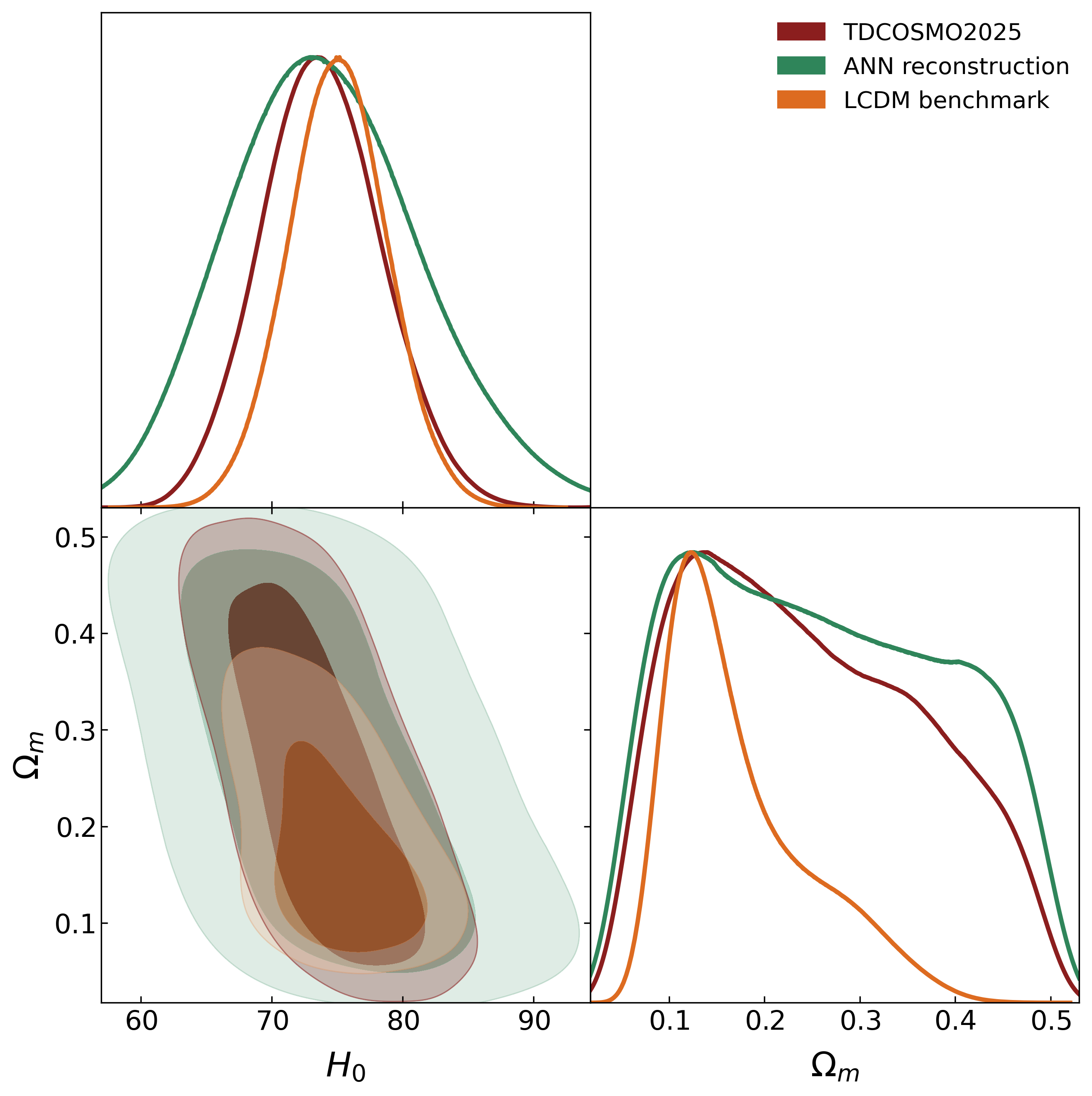}
	\caption{Comparison of the flat-$\Lambda$CDM constraints of $(H_0,\Omega_m)$ obtained from the public TDCOSMO2025 lensing-only posterior, the ANN reconstruction projected using the full effective covariance, and the $\Lambda$CDM benchmark.}
	\label{fig:app_tdcosmo_ann_corner}
\end{figure}

As an additional validation, we project the reconstructed ANN distance vector and its full effective covariance into the flat-$\Lambda$CDM $(H_0,\Omega_m)$ parameter space and compare the result with the public TDCOSMO2025 lensing-only posterior. We also show the $\Lambda$CDM benchmark obtained from the same validation pipeline. Figure~\ref{fig:app_tdcosmo_ann_corner} shows that the ANN constraints overlap with the official lensing-only posterior within the current uncertainties. The comparison is used only to assess the consistency of the reconstructed distance information and does not represent an independent determination of the absolute distance scale.

% =========================================================
\section{Calibration and extended sample-definition tests}
\label{app:calibration_support}
% =========================================================
The reference calibration is restricted to the lens-supported interval
$0.2283\leq z\leq0.745$, and GRBs outside this range are not used to determine the empirical correlation parameters. The Combo boundary scan is listed in Table~\ref{tab:combo_zcut_scan}. The largest changes occur at low $z_{\rm cut}$, where fewer GRBs constrain the calibration, while the results become substantially more stable for $z_{\rm cut}\gtrsim0.65$.
\begin{table}[htbp]
		\centering
		\caption{Calibration-boundary scan for the Combo relation in flat $\Lambda$CDM.}
		\label{tab:combo_zcut_scan}
		\begin{tabular}{cccccc}
			\toprule
			$z_{\rm cut}$ &
			\multicolumn{2}{c}{Fixed $\lambda_C=0$} &
			\multicolumn{3}{c}{Free $\lambda_C$} \\
			\cmidrule(lr){2-3}
			\cmidrule(l){4-6}
			&
			$H_0$ &
			$\Omega_m$ &
			$H_0$ &
			$\Omega_m$ &
			$\lambda_C$ \\
\midrule
0.550 & 65.08 & 0.786 & 65.10 & 0.741 & 0.070 \\
0.600 & 73.06 & 0.738 & 74.27 & 0.713 & 0.019 \\
0.650 & 81.95 & 0.683 & 83.09 & 0.691 & -0.047 \\
0.700 & 82.27 & 0.626 & 83.47 & 0.675 & -0.086 \\
0.745 & 79.07 & 0.622 & 78.71 & 0.635 & -0.083 \\

\bottomrule
		\end{tabular}
	\end{table}
	
We also retain $z_{\rm cut}=1.0$ and $1.25$ as deliberate out-of-support sensitivity tests. They are not used in the reference calibration, but show how the inferred parameters respond when the calibration/cosmology division is moved beyond the direct lens support. For the sample-definition tests we use $\Delta\theta/\sigma_{\theta,\rm base}=(\theta_{\rm test}-\theta_{\rm base})/\sigma_{\theta,\rm base}$, with $\theta\in\{H_0,\Omega_m\}$. Most deterministic sample modifications remain within the baseline statistical uncertainty, and the larger responses arise mainly from redshift removal and resampling, with Combo showing the broader dependence.
	
\begin{figure*}[htbp]
	\centering
	\includegraphics[width=0.85\textwidth]{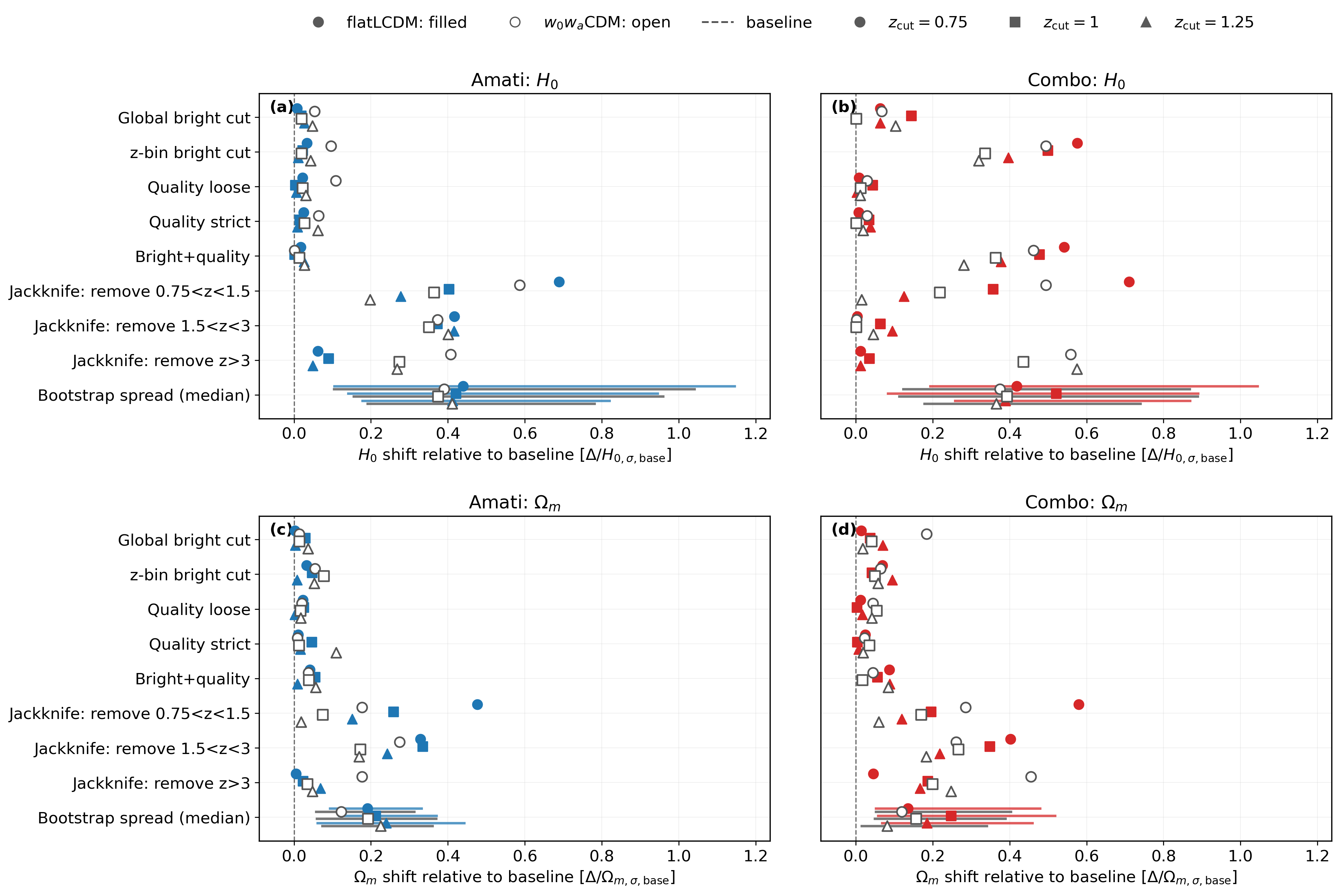}
	\caption{Sample-definition sensitivity tests for $H_0$ and $\Omega_m$. Circles, squares, and triangles denote $z_{\rm cut}=0.75$, $1.0$, and $1.25$, respectively. The latter two cases extend beyond the direct lens support and are shown only as sensitivity tests. Filled and open markers denote flat $\Lambda$CDM and $w_0w_a$CDM. Bootstrap bars show the spread among resampled realizations.}
		\label{fig:app_sampledef_h0_om_multizcut}
\end{figure*}

% =========================================================
\section{Additional GRB inference diagnostics}
\label{app:grb_diagnostics}
% =========================================================
	
\begin{figure*}[htbp]
	\centering
	\includegraphics[width=0.45\textwidth]{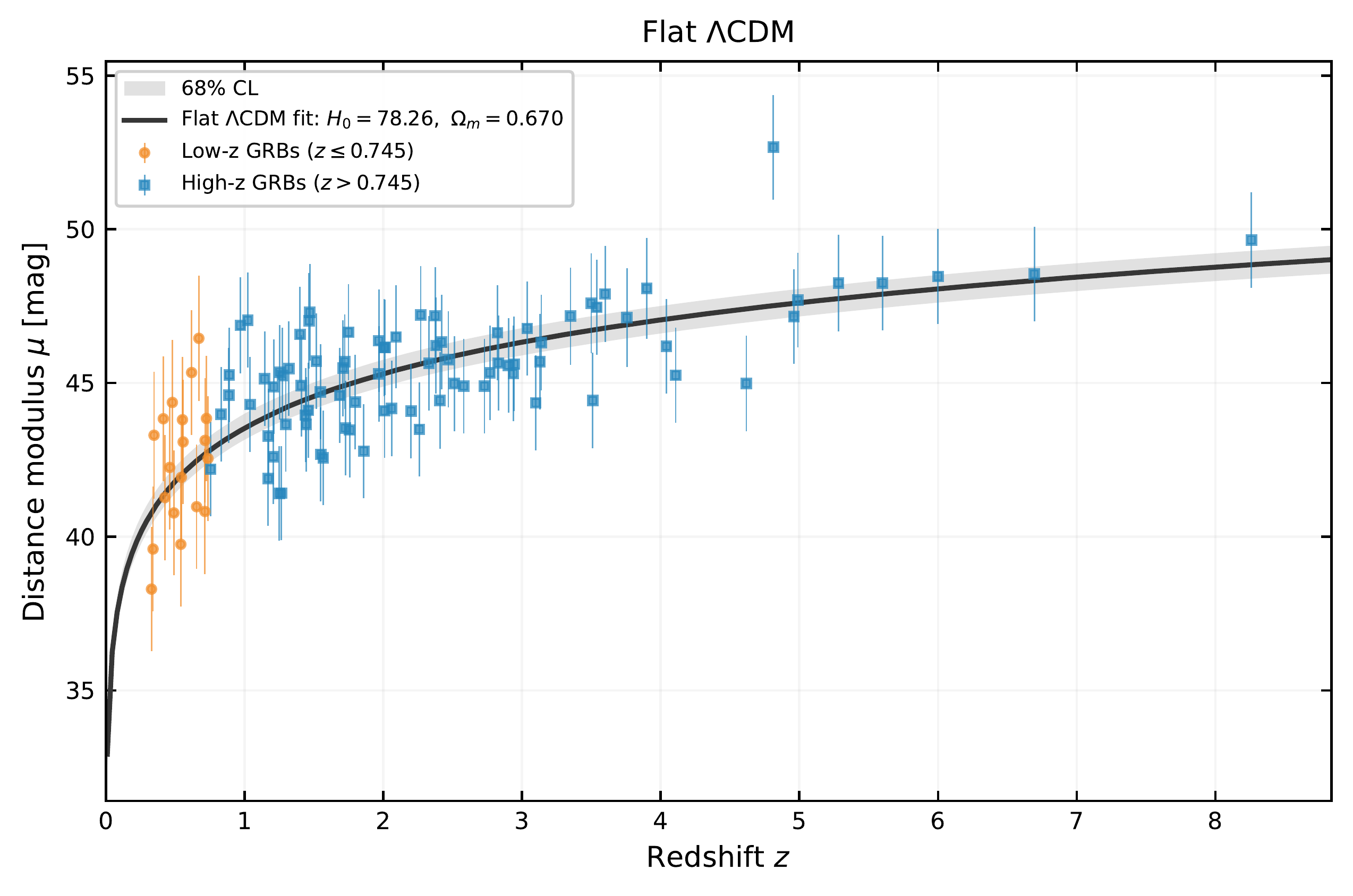}
	\includegraphics[width=0.45\textwidth]{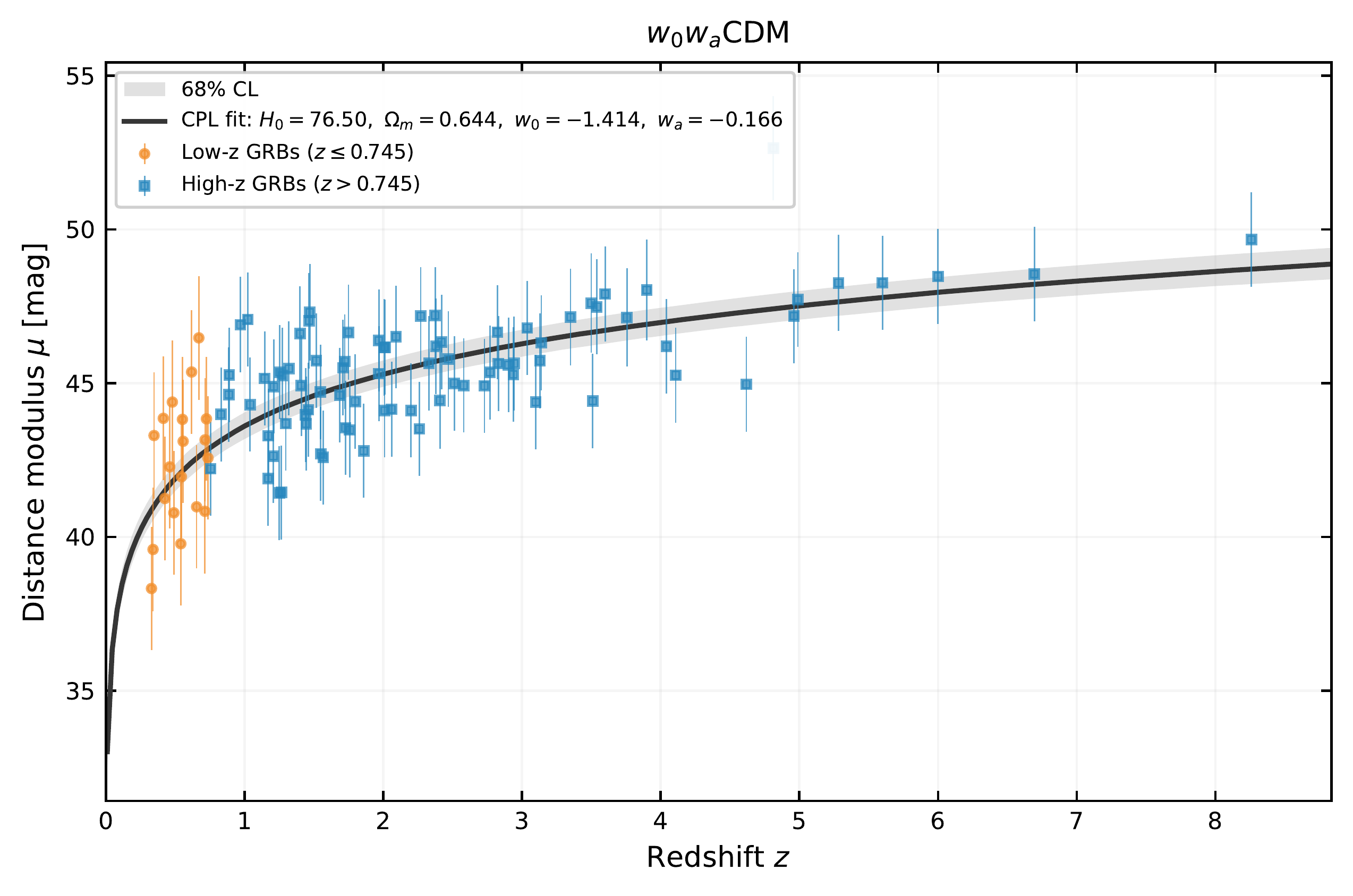}
	\includegraphics[width=0.45\textwidth]{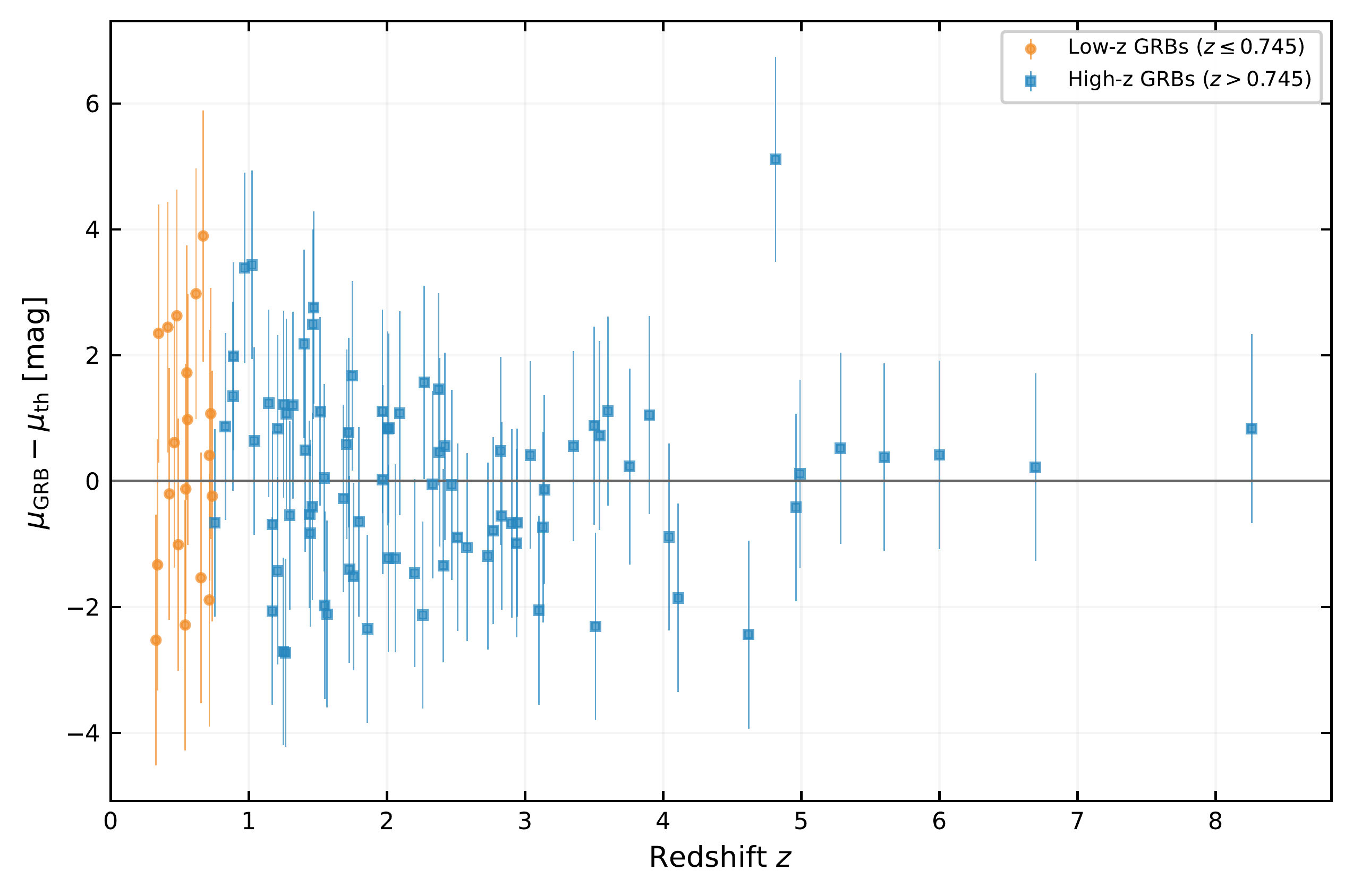}
	\includegraphics[width=0.45\textwidth]{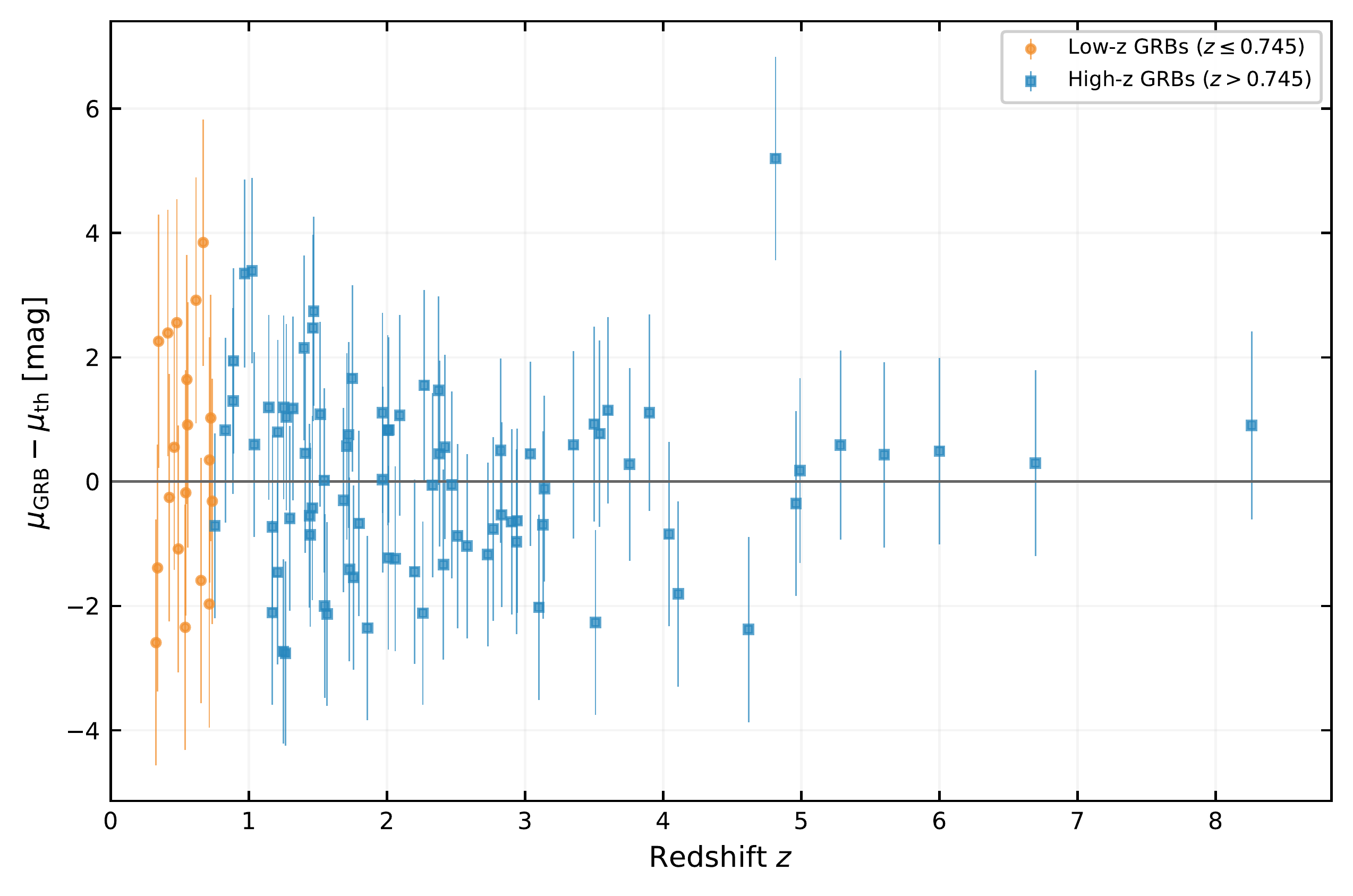}
	\caption{Amati Hubble diagrams and residuals. The left and right columns correspond to flat $\Lambda$CDM and $w_0w_a$CDM, respectively.}
	\label{fig:hubble_amati}
\end{figure*}

\begin{figure*}[htbp]
	\centering
	\includegraphics[width=0.45\textwidth]{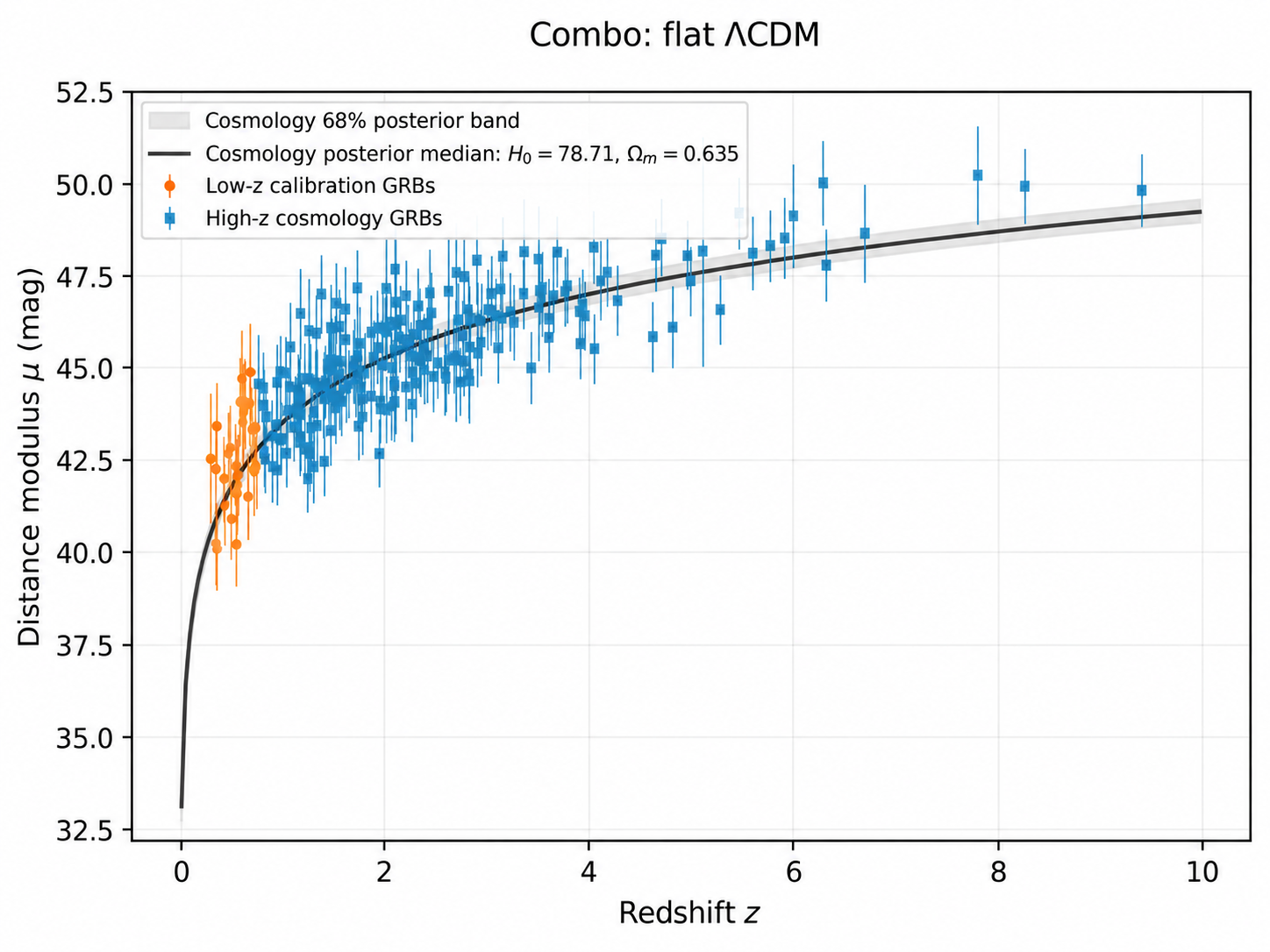}
	\includegraphics[width=0.45\textwidth]{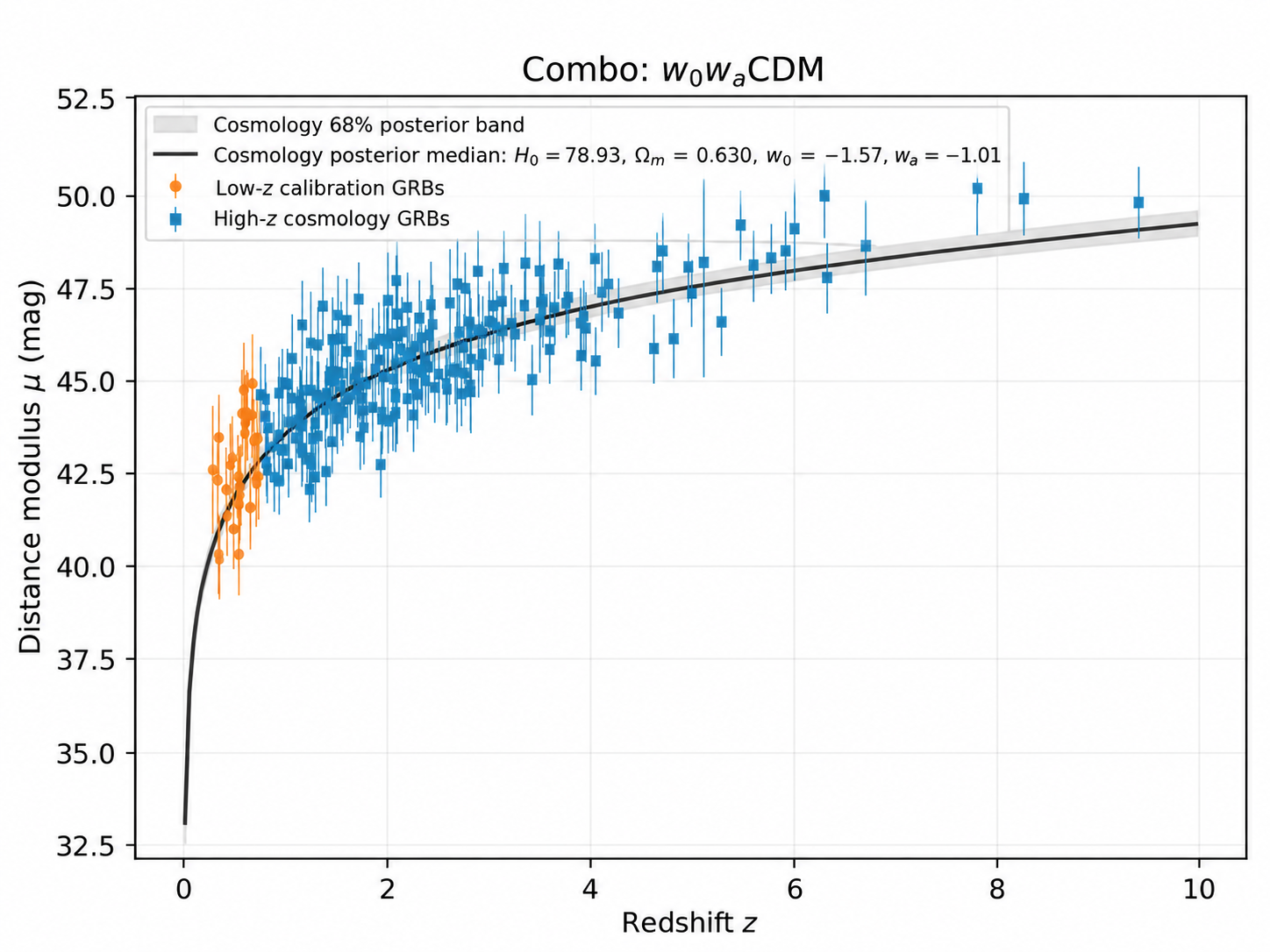}
	\includegraphics[width=0.45\textwidth]{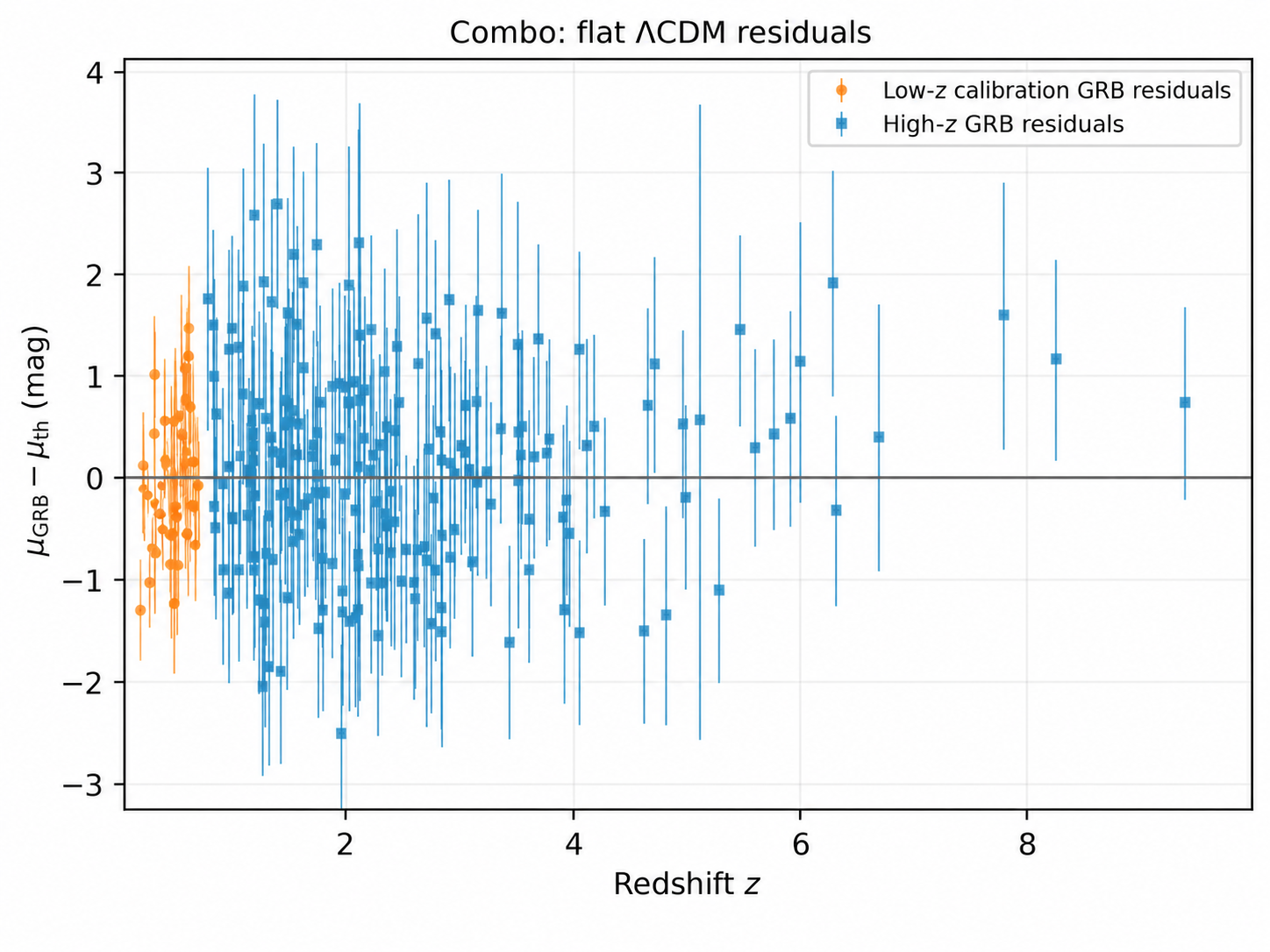}
	\includegraphics[width=0.45\textwidth]{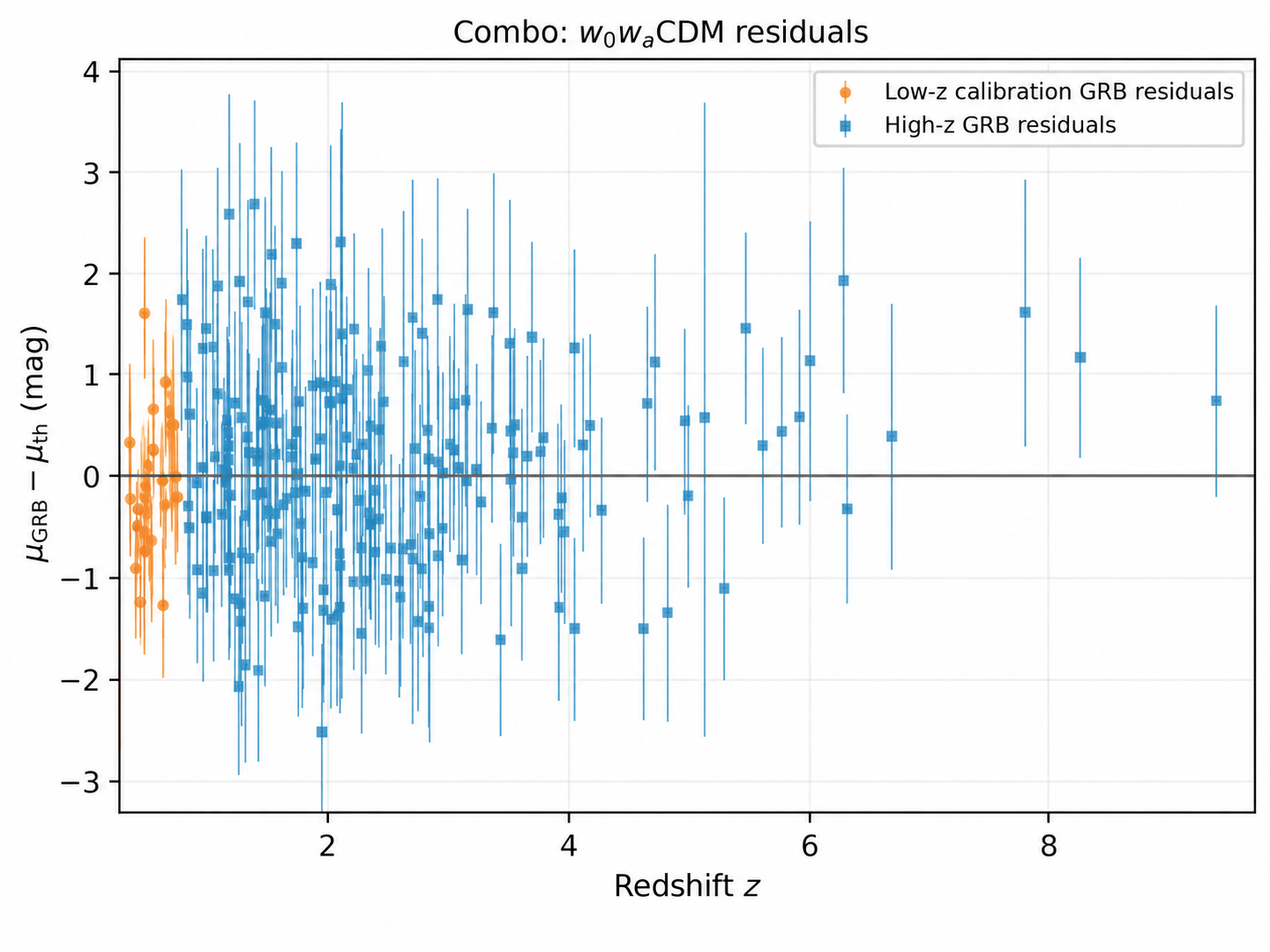}
	\caption{Same as Fig.~\ref{fig:hubble_amati}, but for the Combo relation.}
	\label{fig:hubble_combo}
\end{figure*}

\begin{figure*}[htbp]
	\centering
	\includegraphics[width=0.48\textwidth]{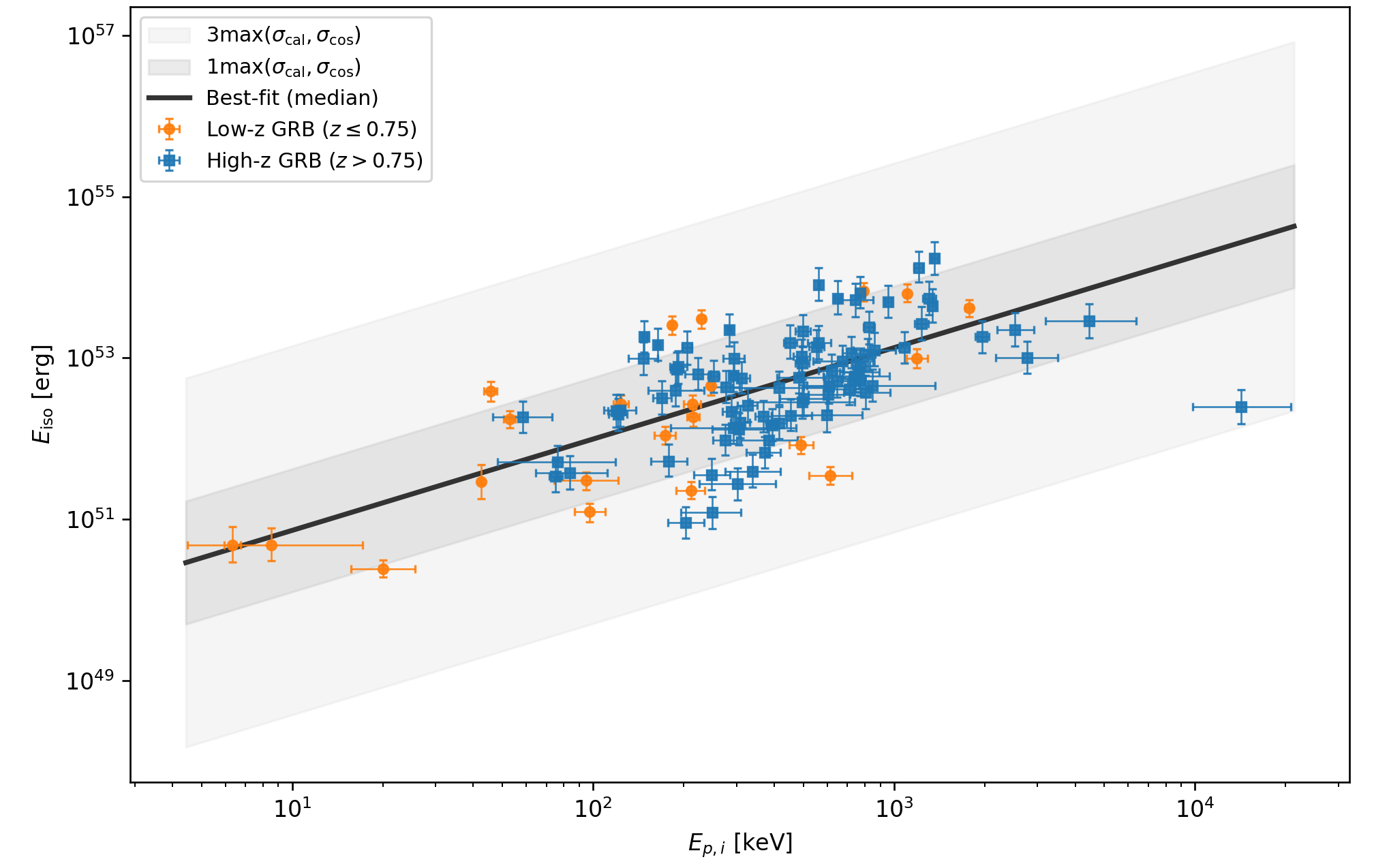}
	\includegraphics[width=0.45\textwidth]{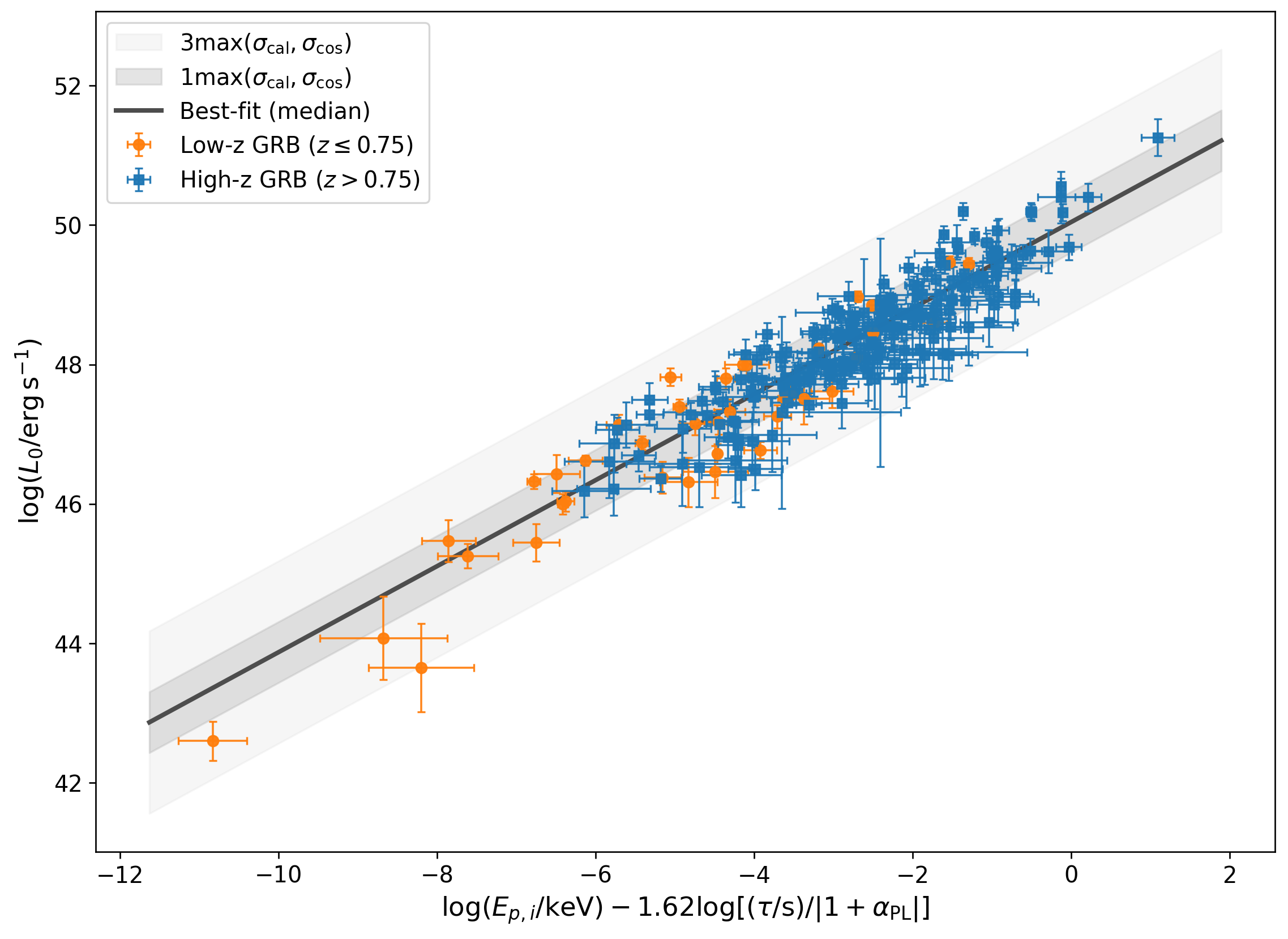}
	\caption{Empirical GRB correlations used in the distance inference: Amati (left) and Combo (right). Calibration and high-redshift cosmological GRBs are shown separately.}
	\label{fig:grb_relations}
\end{figure*}

\begin{figure*}[htbp]
	\centering
	\includegraphics[width=0.45\textwidth]{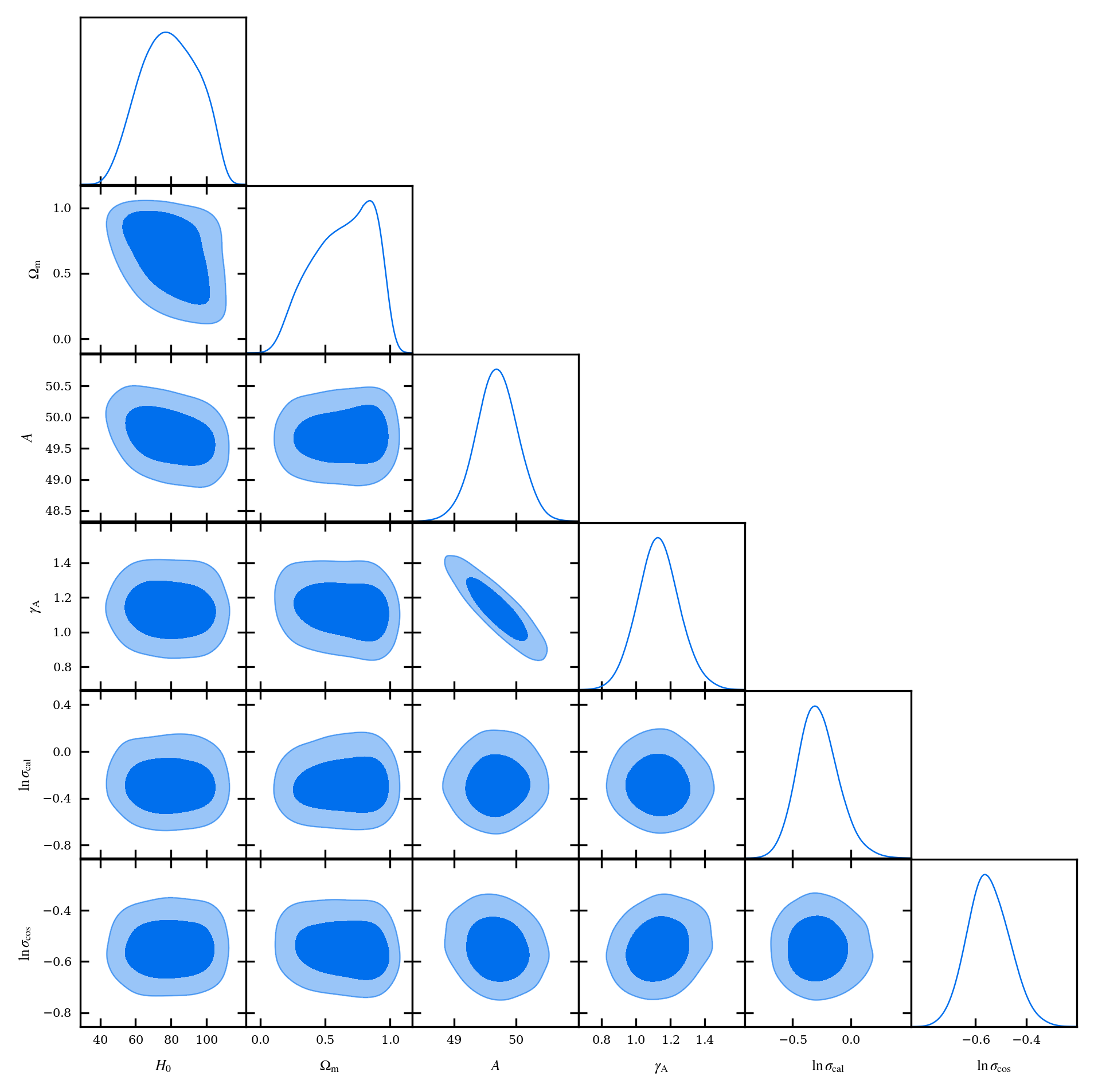}
	\includegraphics[width=0.45\textwidth]{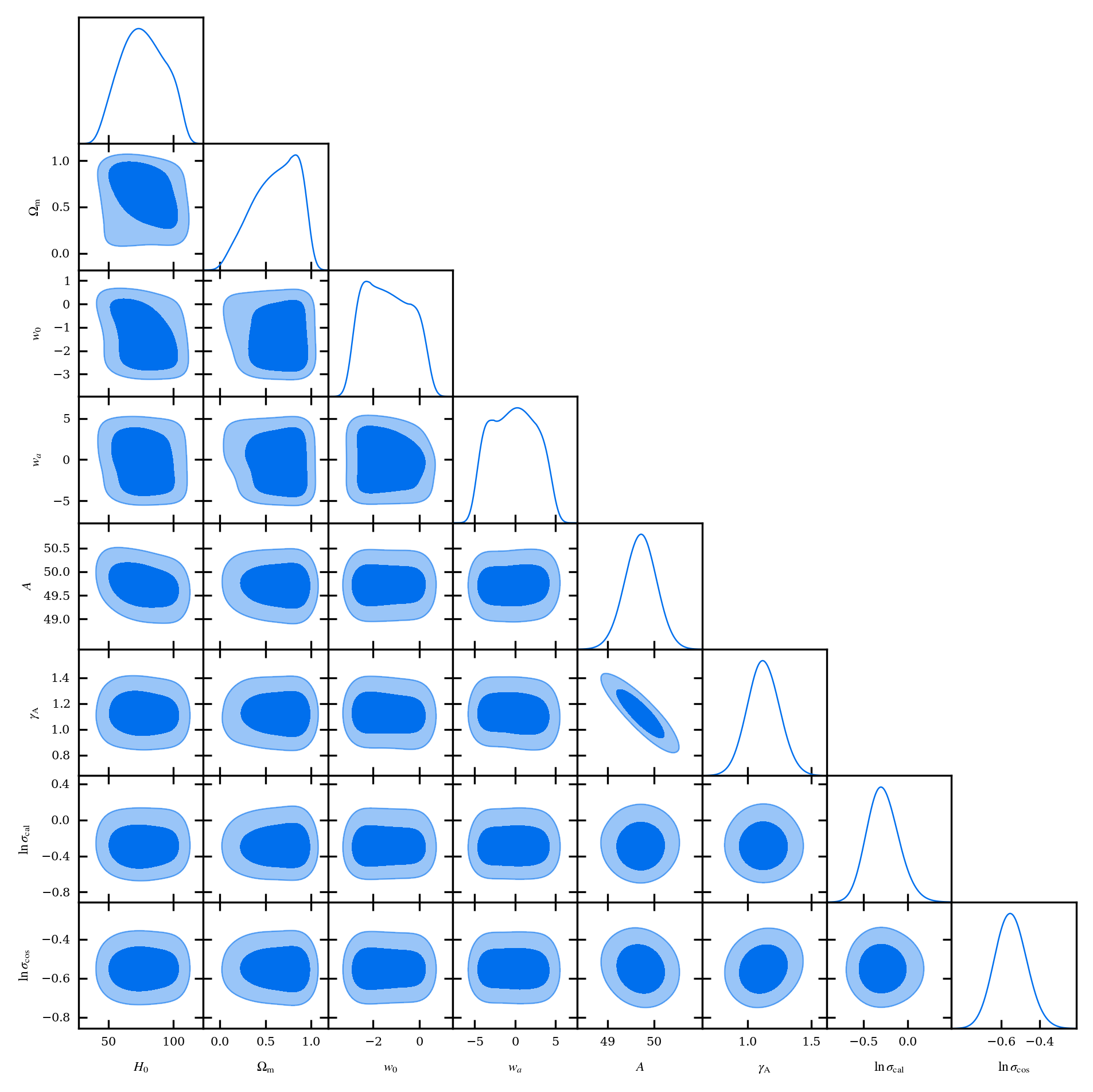}
	\includegraphics[width=0.45\textwidth]{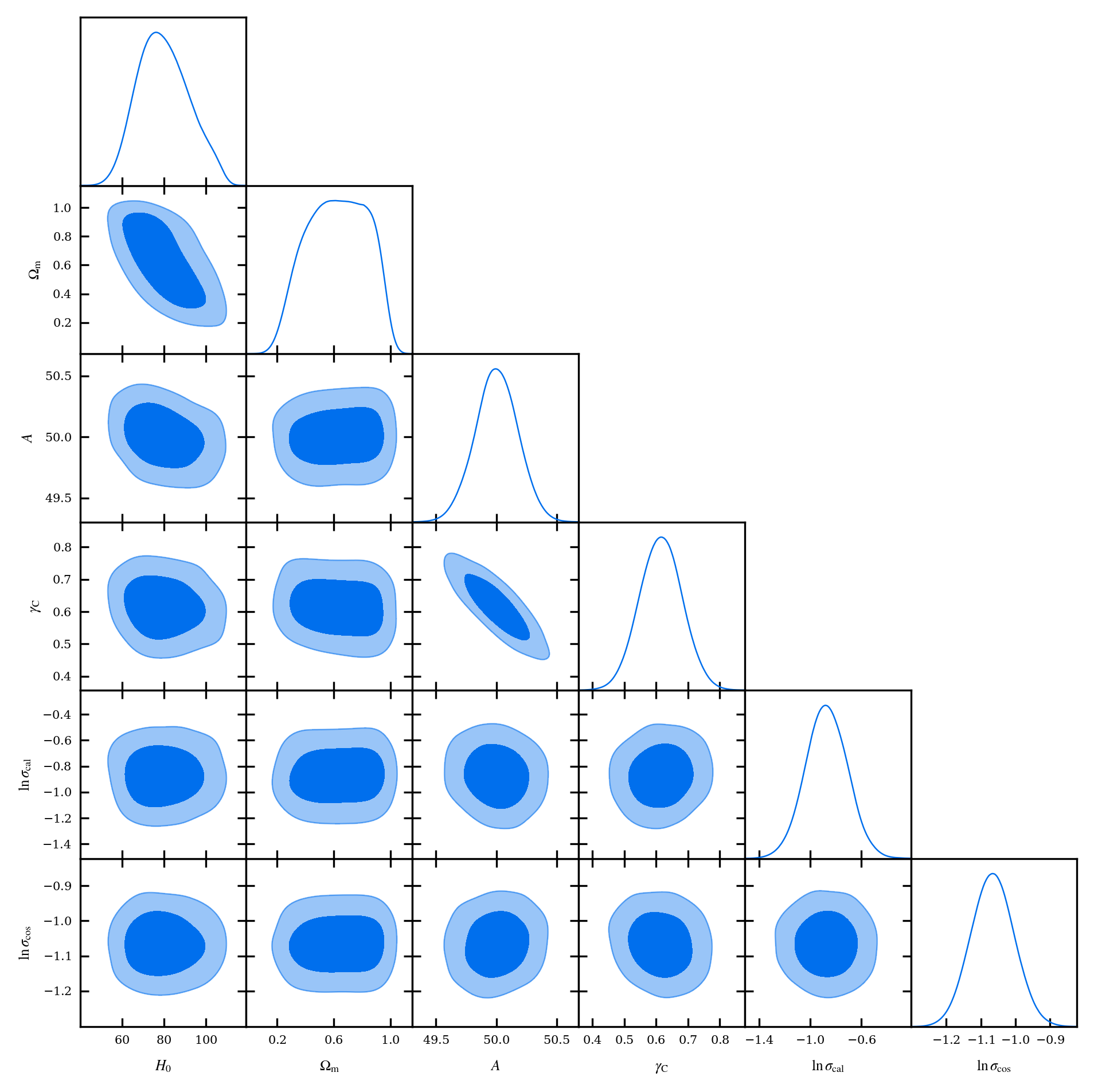}
	\includegraphics[width=0.45\textwidth]{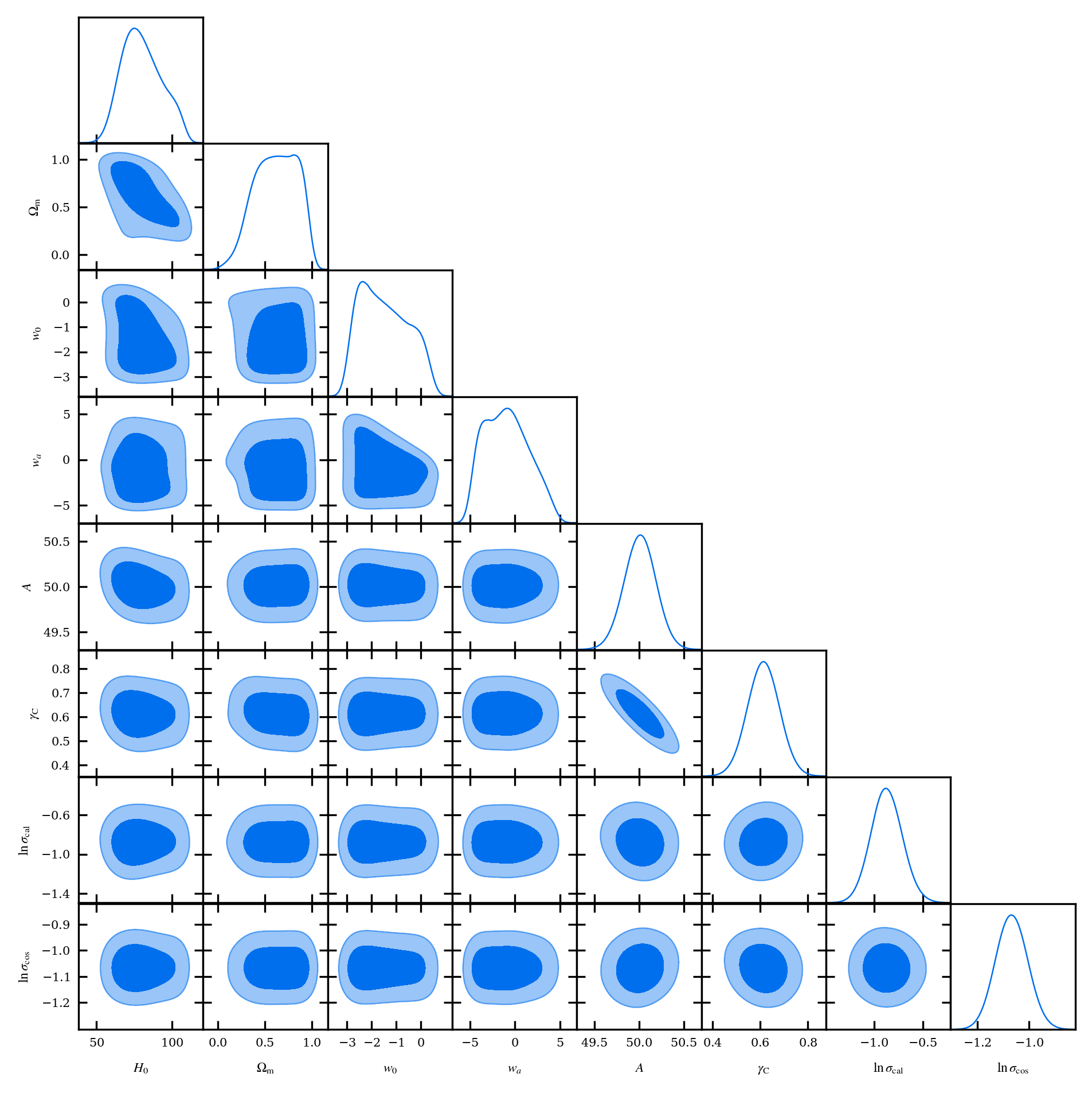}
	\caption{Joint posterior distributions. The upper and lower rows show Amati and Combo, while the left and right columns correspond to flat $\Lambda$CDM and $w_0w_a$CDM.}
		\label{fig:grb_corner_posteriors}
	\end{figure*}

This second collects the Hubble diagrams, residuals, empirical-correlation planes, and joint posterior distributions for the reference analyses. The baseline calibration uses only
$0.2283\leq z\leq0.745$, and the ANN covariance is propagated in the hierarchical likelihood. Figures~\ref{fig:hubble_amati}--\ref{fig:grb_corner_posteriors} provide visual checks of the inferred distances and parameter correlations. And, the broad $w_0w_a$CDM posteriors illustrate the limited dark-energy constraining power of the present GRB samples.
	
Figure~\ref{fig:grb_relations} shows the empirical GRB-correlation planes. Calibration and cosmological GRBs are displayed separately to emphasize that only the former receive direct distance information from the ANN reconstruction.
	
The joint posterior distributions are shown in Fig.~\ref{fig:grb_corner_posteriors}. The cosmological parameters and GRB-correlation parameters are sampled simultaneously, so the displayed correlations include the uncertainty propagated from the calibration into the cosmological inference. The broad $w_0w_a$CDM distributions illustrate the weak constraining power of the present GRB samples on the dark-energy parameters.

% =========================================================
\section{Additional statistical checks and sample properties}
\label{app:additional_checks}
% =========================================================
	
This Appendix collects numerical checks that complement the robustness tests discussed in Sect.~\ref{subsec:results_robustness}. They are separated from the primary results because they do not alter the reference likelihood or the baseline cosmological interpretation.
	
%\subsection{Prior sensitivity and MCMC convergence}
%\label{app:priors_convergence}
	
The reference prior ranges are listed in Sect.~\ref{subsec:cosmo_models}. To test whether the broad cosmological posteriors are significantly affected by the nominal boundaries, we repeat selected analyses with wider ranges $H_0\in[0,140],\Omega_m\in[0,1]$, and, for the widest $w_0w_a$CDM tests $w_0\in[-5,1], w_a\in[-10,10]$. The corresponding shifts are small compared with the posterior widths and do not alter the qualitative conclusions discussed in the main text.
	
The chains are monitored using the integrated autocorrelation time. Production runs are retained only when the number of post-burn-in samples satisfies $N_{\rm post}>50\tau_{\max}$, where $\tau_{\max}$ is the largest integrated autocorrelation time among the sampled parameters.
\begin{table}[htbp]
	\centering
	\scriptsize
	\caption{Convergence diagnostics for the principal production chains.}
	\label{tab:mcmc_convergence}
	\begin{tabular}{lccc}
		\toprule
		Run & $N_{\rm post}$ & $\tau_{\max}$ & $N_{\rm post}/\tau_{\max}$ \\
\midrule
\multicolumn{4}{l}{\textit{Flat $\Lambda$CDM}} \\
Amati, $\lambda_A=0$ & 6000 & 92.35 & 64.97 \\
Amati, free $\lambda_A$ & 13000 & 134.47 & 96.67 \\
Combo, $\lambda_C=0$ & 6000 & 88.46 & 67.83 \\
Combo, free $\lambda_C$ & 13000 & 167.46 & 77.63 \\
\addlinespace
\multicolumn{4}{l}{\textit{$w_0w_a$CDM}} \\
Amati, $\lambda_A=0$ & 31000 & 254.42 & 121.84 \\
Combo, $\lambda_C=0$ & 31000 & 480.04 & 64.58 \\
		\bottomrule
	\end{tabular}
\end{table}

%\subsection{Calibration-transfer test}
%\label{app:calibration_transfer}
	
To determine whether the high-$\Omega_m$ Combo solution at $\lambda_C=0$ is produced by reverse updating of the low-redshift calibration, we repeat the high-redshift fit after transferring calibration-only constraints on the relation parameters. The persistence of the high-$\Omega_m$ solution in Table~\ref{tab:calibration_transfer} shows that this effect is not primarily caused by reverse updating of the calibration relation.
\begin{table}[htbp]
	\centering
	\scriptsize
	\caption{Combo calibration-transfer tests in flat $\Lambda$CDM. Values are posterior medians with central $68\%$ intervals.}
	\label{tab:calibration_transfer}
	\begin{tabular}{lcccc}
		\toprule
		Calibration treatment & $\lambda_C$ treatment & $H_0$ & $\Omega_m$ & $\lambda_C$ \\
		\midrule
Full joint fit & fixed & $80.97^{+11.73}_{-10.04}$ & $0.811^{+0.129}_{-0.194}$ & 0 \\
Gaussian calibration transfer & fixed & $81.20^{+12.00}_{-10.31}$ & $0.806^{+0.132}_{-0.200}$ & 0 \\
Tight calibration transfer & fixed & $84.35^{+7.50}_{-5.23}$ & $0.830^{+0.116}_{-0.187}$ & 0 \\
Full joint fit & free & $79.92^{+13.80}_{-11.43}$ & $0.668^{+0.226}_{-0.287}$ & $0.319^{+0.216}_{-0.201}$ \\
Gaussian calibration transfer & free & $80.08^{+13.68}_{-11.39}$ & $0.668^{+0.225}_{-0.284}$ & $0.321^{+0.214}_{-0.203}$ \\
Tight calibration transfer & free & $80.92^{+11.48}_{-8.00}$ & $0.667^{+0.226}_{-0.285}$ & $0.411^{+0.205}_{-0.195}$ \\
\bottomrule
	\end{tabular}
\end{table}
	
%\subsection{Overlap between the Amati and Combo samples}
%\label{app:sample_overlap}
	
The Amati sample is fully contained in the Combo catalogue: all 19 calibration GRBs and all 84 cosmological GRBs are common to both, giving an overall overlap of $100\%$ for the Amati analysis sample. The two cosmological analyses are therefore not statistically independent. The overlap is summarized in Table~\ref{tab:sample_overlap}.
	
\begin{table}[htbp]
		\centering
		\caption{
			Overlap between the Amati and Combo samples.
		}
		\label{tab:sample_overlap}
		\begin{tabular}{lccc}
			\toprule
			Subset &
			$N_A$ &
			$N_C$ &
			$N_{A\cap C}$ \\
			\midrule
			Calibration sample & 19 & 32 & 19 \\
			Cosmological sample & 84 & 205 & 84 \\
			Full analysis sample & 103 & 237 & 103 \\
			
			\bottomrule
		\end{tabular}
	\end{table}
	
The robustness tests give a consistent picture in both flat $\Lambda$CDM and $w_0w_a$CDM. We do not repeat here the numerical results already reported in the preceding sections. The reference and free-redshift-term constraints are given in Tables~\ref{tab:cosmo_results} and \ref{tab:evolution_results}, while the fixed-$\lambda$ dependence is shown in Figs.~\ref{fig:layer2_sensitivity_flatlcdm} and \ref{fig:layer2_sensitivity_cpl}. The calibration-transfer results, calibration-boundary scan, and broader sample-definition tests are reported in Table~\ref{tab:calibration_transfer}, Table~\ref{tab:combo_zcut_scan}, and Fig.~\ref{fig:app_sampledef_h0_om_multizcut}, respectively. The common-scatter and wider-prior tests in flat $\Lambda$CDM are discussed in Sect.~\ref{subsec:results_robustness}, while the injection--recovery analysis is presented in Sect.~\ref{subsec:lambda_results}.

For the archived samples, we quantify the response to each variation through
\begin{equation}
	\frac{\Delta\theta}{\sigma_{\theta,\rm base}}=\frac{\left|\theta_{\rm test}-\theta_{\rm base}\right|}{\sigma_{\theta,\rm base}},
	\quad
	\theta\in\{H_0,\Omega_m,w_0,w_a\},
	\label{eq:comparative_normalized_shift}
\end{equation}
where $\sigma_{\theta,\rm base}$ is the symmetrized $68\%$ uncertainty of the corresponding baseline run. These archived calculations are used only to quantify sensitivity to sample composition and are not adopted as reference cosmological constraints. In all such tests the ANN calibration interval remains fixed at $0.2283\leq z\leq0.745$.

Table~\ref{tab:comparative_robustness} collects only the supplementary results that are useful for a direct comparison between the two cosmological models and are not already tabulated in detail elsewhere. For the sample-definition tests, only the largest deterministic normalized shift in each relation and cosmological model is listed.
\begin{table}[htbp]
	\centering
	\scriptsize
	\setlength{\tabcolsep}{2.8pt}
	\renewcommand{\arraystretch}{1.06}
	\caption{Selected robustness diagnostics. Sample-definition rows report the largest normalized posterior shift among the deterministic sample-selection tests. Wide-prior rows summarize the corresponding $w_0w_a$CDM stress tests.
	}
	\label{tab:comparative_robustness}
	\begin{tabular}{@{}llp{2.35cm}p{3.15cm}@{}}
		\toprule
		Model & Relation & Test & Response \\
		\midrule
		\multicolumn{4}{l}{\textit{Sample-definition sensitivity}} \\
		\addlinespace[1pt]
		$\Lambda$CDM &Amati &Remove $0.75\leq z<1.5$ &$\Delta H_0/\sigma=0.643$ \\
		$\Lambda$CDM &Combo &Remove $0.75\leq z<1.5$ &$\Delta H_0/\sigma=0.769$ \\
		$w_0w_a$CDM &Amati &Remove $0.75\leq z<1.5$ &$\Delta H_0/\sigma=0.586$ \\
		$w_0w_a$CDM &Combo &Remove $z\geq3$ &$\Delta w_a/\sigma=0.782$ \\
		\addlinespace
		\multicolumn{4}{l}{\textit{Wide-prior $w_0w_a$CDM stress tests}} \\
		\addlinespace[1pt]
		$w_0w_a$CDM &Amati &$\lambda_A=0$ &
		$\begin{array}{l}
			w_0=-2.595^{+2.144}_{-1.671},\\[-1pt]
			w_a=-1.459^{+6.207}_{-5.851}
		\end{array}$ \\
		$w_0w_a$CDM &Combo &free $\lambda_C$ &
		$\begin{array}{l}
			w_0=-2.502^{+2.257}_{-1.766},\\[-1pt]
			w_a=-1.662^{+5.529}_{-5.717}
		\end{array}$ \\
		\bottomrule
	\end{tabular}
	\vspace{0.25em}
	\begin{minipage}{0.96\columnwidth}
		\footnotesize
		\textit{Notes.}
		The normalized shifts are defined relative to the marginalized posterior uncertainty of the corresponding baseline analysis. The archived sample-definition branches are not identical event by event and are used only for response comparisons. The wide priors are
		$H_0\in[0,140],\Omega_m\in[0,1]$, $w_0\in[-5,1]$, and $w_a\in[-10,10]$.
	\end{minipage}
\end{table}

In flat $\Lambda$CDM, changes in the intrinsic-scatter prescription and cosmological prior ranges produce shifts smaller than the current posterior widths. The larger responses are associated with changes in the redshift distribution and sample composition. The additional Combo-slope tests likewise do not remove the stronger dependence of the Combo constraints on the phenomenological redshift term. Furthermore, the calibration-transfer analysis shows that the high-$\Omega_m$ solution obtained for $\lambda_C=0$ is not mainly produced by reverse updating of the low-redshift calibration. The injection--recovery tests provide a complementary check of the same redshift--cosmology degeneracy.

The $w_0w_a$CDM sample-definition tests exhibit a similar overall pattern to those in flat $\Lambda$CDM. For all deterministic sample-selection variations considered, the shifts in the four cosmological parameters remain below their corresponding baseline posterior uncertainties. For Amati, the largest normalized shift occurs in $H_0$ after excluding $0.75\leq z<1.5$, with $|\Delta H_0|/\sigma_{H_0,\rm base}=0.586$. For Combo, the largest response is found in $w_a$ after excluding the $z\geq3$ events, yielding $|\Delta w_a|/\sigma_{w_a,\rm base}=0.782$. These results indicate that extending the background cosmology to a time-dependent dark-energy equation of state does not substantially increase the sensitivity to the sample-selection variations considered here. The different redshift intervals associated with the largest shifts suggest that the Amati and Combo relations respond differently to changes in sample composition, with the largest Combo response arising from the high-redshift subset.

Bootstrap resampling also gives relatively small typical shifts, although individual realizations can reach order-unity deviations. The wide-prior $w_0w_a$CDM runs broaden the dark-energy posteriors substantially, as expected from the limited constraining power of the current GRB samples, but do not isolate a departure from $(w_0,w_a)=(-1,0)$. The $w_0w_a$CDM analysis is therefore used here to examine the dependence of the inferred constraints on the assumed expansion history rather than as evidence for dynamical dark energy.

\end{appendix}

\end{document}